\documentstyle[a4,12pt]{report}
\begin{document}

\newcommand{\ds}{\displaystyle}

\newcommand{\th}{\theta}
\newcommand{\tb}{\bar{\theta}}
\newcommand{\TH}{\Theta}
\newcommand{\TB}{\bar{\Theta}}
\newcommand{\la}{\lambda}
\newcommand{\LA}{\Lambda}
\newcommand{\LB}{\bar{\LA}}
\newcommand{\OM}{\Omega}
\newcommand{\DE}{\Delta}
\newcommand{\bt}{\bar{\tau}}
\newcommand{\zb}{\bar z}
\newcommand{\pa}{\partial}
\newcommand{\dab}{\bar D}
\newcommand{\ZB}{\bar Z}
\newcommand{\pab}{ \bar{\partial} }
\newcommand{\HT}{ {H_{\th}}^z }
\newcommand{\HB}{ {H_{\tb}}^z }
\newcommand{\HO}{ H_{\th} ^{\ \th} }
\newcommand{\HZ}{ H_{\zb} ^{\ \th} }
\newcommand{\HZB}{ H_{\zb} ^{\ z} }
\newcommand{\HOB}{ H_{\tb} ^{\ \th} }

\newcommand{\zt}{\tilde z}
\newcommand{\dt}{\tilde D}
\newcommand{\pabt}{\tilde{\pab}}
\newcommand{\dabt}{\tilde{\dab}}

\newcommand{\thm}{\theta^-}
\newcommand{\tbm}{{\bar{\theta}}^-}
\newcommand{\zz}{\cal Z}

\newcommand{\pr}{\prime}

\newcommand{\hzt}{ h_z^{\ \th} }
\newcommand{\hzbt}{ h_{\zb}^{\ \th} }
\newcommand{\hztb}{ h_z^{\ \tb} }
\newcommand{\hzbtb}{ h_{\zb}^{\ \tb} }
\newcommand{\htt}{ h_{\th}^{\ \th} }
\newcommand{\httb}{ h_{\th}^{\ \tb} }
\newcommand{\htbt}{ h_{\tb}^{\ \th} }
\newcommand{\htbtb}{ h_{\tb}^{\ \tb} }
\newcommand{\htz}{ h_{\th}^{\ z} }
\newcommand{\htzb}{ h_{\th}^{\ \zb} }
\newcommand{\htbz}{ h_{\tb}^{\ z } }
\newcommand{\htbzb}{ h_{\tb}^{\ \zb} }

\newcommand{\LAm}{\LA ^-}
\newcommand{\LBm}{\LB ^-}
\newcommand{\taub}{\bar{\tau}}
\newcommand{\taum}{\tau ^-}
\newcommand{\taubm}{\taub ^-}

\newcommand{\Hzz}{ H_{z}^{\ z} }
\newcommand{\Hzt}{ H_{z}^{\ \th} }
\newcommand{\Hztb}{ H_{z}^{\ \tb} }
\newcommand{\Htz}{ H_{\th}^{\ z} }
\newcommand{\Htt}{ H_{\th}^{\ \th} }
\newcommand{\Httb}{ H_{\th}^{\ \tb} }
\newcommand{\Htbt}{ H_{\tb}^{\ \th} }
\newcommand{\Htbtb}{ H_{\tb}^{\ \tb} }
\newcommand{\Htbz}{ H_{\tb}^{\ z } }

\newcommand{\Hzzb}{ H_{z}^{\ \zb} }
\newcommand{\Hztm}{ H_{z}^{\ \th ^-} }
\newcommand{\Hztbm}{ H_{z}^{\ \tb ^-} }
\newcommand{\Htzb}{ H_{\th}^{\ \zb} }
\newcommand{\Httm}{ H_{\th}^{\ \th ^-} }
\newcommand{\Httbm}{ H_{\th}^{\ \tb ^-} }
\newcommand{\Htbtm}{ H_{\tb}^{\ \th ^-} }
\newcommand{\Htbtbm}{ H_{\tb}^{\ \tb ^-} }
\newcommand{\Htbzb}{ H_{\tb}^{\ \zb } }

\newcommand{\Hzbz}{ H_{\zb}^{\ z} }
\newcommand{\Hzbt}{ H_{\zb}^{\ \th} }
\newcommand{\Hzbtb}{ H_{\zb}^{\ \tb} }
\newcommand{\Htmz}{ H_{\th ^-}^{\ z} }
\newcommand{\Htmt}{ H_{\th ^-}^{\ \th} }
\newcommand{\Htmtb}{ H_{\th ^-}^{\ \tb} }
\newcommand{\Htbmt}{ H_{\tb ^-}^{\ \th} }
\newcommand{\Htbmtb}{ H_{\tb ^-}^{\ \tb} }
\newcommand{\Htbmz}{ H_{\tb ^-}^{\ z } }

\newcommand{\Hzbzb}{ H_{\zb}^{\ \zb} }
\newcommand{\Hzbtm}{ H_{\zb}^{\ \th ^-} }
\newcommand{\Hzbtbm}{ H_{\zb}^{\ \tb ^-} }
\newcommand{\Htmzb}{ H_{\th ^-}^{\ \zb} }
\newcommand{\Htmtm}{ H_{\th ^-}^{\ \th ^-} }
\newcommand{\Htmtbm}{ H_{\th ^-}^{\ \tb ^-} }
\newcommand{\Htbmtm}{ H_{\tb ^-}^{\ \th ^-} }
\newcommand{\Htbmtbm}{ H_{\tb ^-}^{\ \tb ^-} }
\newcommand{\Htbmzb}{ H_{\tb ^-}^{\ \zb } }

\newcommand{\htmt}{ h_{\thm}^{\ \th} }
\newcommand{\httm}{ h_{\th}^{\ \thm} }
\newcommand{\htmtm}{ h_{\thm}^{\ \thm} }
\newcommand{\htmtb}{ h_{\thm}^{\ \tb} }
\newcommand{\httbm}{ h_{\th}^{\ \tbm} }
\newcommand{\htmtbm}{ h_{\thm}^{\ \tbm} }
\newcommand{\htbmtm}{ h_{\tbm}^{\ \thm} }
\newcommand{\htbmt}{ h_{\tbm}^{\ \th} }
\newcommand{\htbtm}{ h_{\tb}^{\ \thm} }
\newcommand{\htbmtb}{ h_{\tbm}^{\ \tb} }
\newcommand{\htbmtbm}{ h_{\tbm}^{\ \tbm} }
\newcommand{\htbtbm}{ h_{\tb}^{\ \tbm} }
\newcommand{\htmz}{ h_{\thm}^{\ z} }
\newcommand{\htmzb}{ h_{\thm}^{\ \zb} }

\newcommand{\mb}{ \bar{\mu} }
\newcommand{\lb}{ \bar{\la}}

\thispagestyle{empty}

\hfill  LYCEN/9620
\vskip 0.07truecm
\hfill    ENSLAPP-L-616/96
\vskip 0.07truecm
\hfill  hep-th/9609xxx
\vskip 0.09truecm
\hfill  September 1996
\vskip 0.01truecm
\bigskip
\begin{center}
{\bf \Huge{$d=2, \  N=2$}}
\end{center}
\begin{center}
{\bf \Huge{Superconformal Symmetries and Models}}
\end{center}
\vskip 1.3truecm
\centerline{ {\bf Fran\c cois Delduc} $^a$ $\ $ , $\ $
{\bf Fran\c cois Gieres} $^b$  $\ $ , $\ $
{\bf St\'ephane Gourmelen} $^b$ }
\bigskip
\centerline{$^a${\it Laboratoire de Physique Th\'eorique} $^{\S}$}
\centerline{\it ENS Lyon}
\centerline{\it 46, all\'ee d'Italie}
\centerline{\it F - 69364 - Lyon C\'edex 07}
\bigskip
\bigskip
\centerline{$^b${\it Institut de Physique Nucl\'eaire}}
\centerline{\it Universit\'e Claude Bernard (Lyon 1)}
\centerline{\it 43, boulevard du 11 novembre 1918}
\centerline{\it F - 69622 - Villeurbanne C\'edex}
\vskip 1.0truecm
\bigskip
\bigskip
\vskip 1.3truecm

\nopagebreak
We discuss the following aspects
of two-dimensional $N=2$ super\-symmetric theories
defined
on compact super
Riemann surfaces: parametrization of $(2,0)$ and $(2,2)$
superconformal structures in
terms of Beltrami coefficients and formulation of
superconformal models on such surfaces (invariant actions, anomalies and
compensating actions, Ward identities).

\bigskip




\newpage

\chapter{Introduction}

The reasons for studying 2-dimensional $N=2$ superconformal field
theories are numerous and well known (e.g. see \cite{nw}): the
areas of application include string theory, mirror symmetry,
topological field theories, exactly solvable models, quantum
and $W$-gravity.
Since holomorphic factorization represents
a fundamental property of many of these models \cite{fms},
it is particularly
interesting to have a
field theoretic approach in which
holomorphic factorization is realized in a manifest way by virtue of
an appropriate parametrization of the basic variables.

The goal of the present work is to develop such an
approach to the superspace formulation
of (2,2) and (2,0) superconformal models.
In order to describe this approach and its relationship
to other formulations in more detail,
it is useful to summarize briefly previous
work in this field.

The $d=2, N=2$ superconformally invariant coupling of matter fields
to gravity was first discussed in the context of the fermionic
string \cite{ade, bris}. Later on, the analogous
(2,0) supersymmetric theory has been introduced and
sigma-model couplings have been investigated \cite{hw,
bmg, evo}. Some of
this work has been done in component field formalism,
some other in superspace formalism.
The latter has the advantage
that {\em supersymmetry is manifestly realized}
and that {\em field-dependent symmetry algebras are avoided}.
(Such algebras usually occur in the component field formalism (WZ-gauge)
\cite{geo}.)

The geometry of $d=2, N=2$ superspace and the classification
of irreducible multiplets has been analyzed by the authors of
references \cite{ghr, hp, glo, ggw}.
As is well known \cite{ggrs, pcw}, the
quantization of supergravity in superspace requires the explicit
solution of the constraints imposed on the geometry
in terms of prepotential superfields.
In two dimensions, these prepotentials (parametrizing superconformal
classes of metrics) represent superspace expressions of the
Beltrami differentials \cite{gg}. The determination of an
explicit solution
for the (2,0) and (2,2) constraints has been studied
in references \cite{eo, xu, kl, l} and \cite{gz, ot, gw},
respectively.

On the other hand, a field theoretic approach to (ordinary)
conformal models in which holomorphic factorization is
manifestly realized
was initiated by R.Stora and developed
by several authors
\cite{ls, cb}.
This formalism comes in two versions.
One may formulate the theory on a Riemannian manifold
in which case one has to deal with Weyl rescalings of the
metric and with conformal classes of metrics parametrized
by Beltrami coefficients. Alternatively, one may work
on a Riemann surface in which case one simply deals with
complex structures which are equivalent to conformal
classes of metrics.
This Riemannian surface approach
enjoys the following properties.
{\em Locality} is properly taken into account, {\em holomorphic
factorization} is realized manifestly due to a judicious
choice of variables and the theory is {\em globally defined} on a
compact Riemann surface of arbitrary genus. Furthermore,
the fact of working right away on a
Riemann surface (i.e. with a conformal class of metrics)
renders this approach more
{\em economical} since there is no need for introducing
Weyl rescalings and eliminating these degrees of freedom in the
sequel.

The Riemannian manifold approach \cite{cb} has been generalized to
the $N=1$ supersymmetric case in reference
\cite{bbg} and to the $(2,2)$ and $(2,0)$ supersymmetric cases
in references \cite{ot} and \cite{kl}, respectively.
The Riemannian surface approach \cite{ls} has
been extended to the $N=1$ supersymmetric theory in reference \cite{dg}
and was used to prove the
superholomorphic factorization theorem for partition functions on Riemann
surfaces \cite{agn}.
Both of these approaches to superconformal models
are formulated in terms of Beltrami
superfields
(`prepotentials') and their relationship
with the usual (Siegel-Gates like) solution of supergravity constraints
has been discussed in references \cite{dg} and \cite{gg}.
We will come back to this issue in the concluding section where
we also mention further applications. It should be noted
that the generalization to $N=2$ supersymmetry is more subtle
than the one to the $N=1$ theory
due to the appearance of an extra U(1)-symmetry.

Our paper is organized as follows.
We first consider the (2,0) theory since
it allows for simpler
notation and calculations.
Many results for the
$z$-sector of the (2,0) theory have the same form
as those of the $z$-sector of the (2,2) theory
(the corresponding
results for the $\zb$-sector being obtained by complex conjugation).
After a detailed presentation of the (2,0) theory, we simply
summarize the results for the (2,2) theory. Comparison
of our results with those of other approaches will be made within
the text and in
the concluding section.

\chapter{$N=2$ Superconformal symmetry}

In this chapter, we introduce $N=2$ superconformal transformations
and some related notions
\cite{f, cr, jc, bmg, pcw}.
To keep supersymmetry manifest,
all considerations will be carried out in superspace
\cite{wb, ggrs, pcw, geo}, but the
projection of the results to ordinary space will be outlined
in the end.

\section{Superconformal transformations and SRS's}

\subsubsection{Notation and basic relations}

An $N=2$ super Riemann surface (SRS)
is locally parametrized by coordinates
\begin{equation}
\label{coo}
( {\zz} ; {\bar{{\cal Z}}} ) \equiv (z, \th, \tb ;
\zb , \thm, \tbm ) \equiv
(x^{++} , \th^+ , \tb^+ ;
x^{--} , \thm, \tbm )
\ \ ,
\end{equation}
with $z, \zb$ even and $\th, \tb, \thm, \tbm$ odd.
The variables are complex and related by complex conjugation (denoted
by $\ast$):
\[
z^{\ast} = \zb \qquad , \qquad
(\th^+ )^{\ast} = \th ^- \qquad , \qquad
(\tb^+ )^{\ast} = \tb ^-
\ \ .
\]
As indicated in (\ref{coo}),
we will omit the plus-indices of $\th^+$ and $\tb^+$ to simplify
the notation.

The canonical basis of the
tangent space is defined by
$( \pa , \,
D  , \,
\dab ; \,
\pab , \,
D_-  , \,
\dab _- )$  with
\begin{eqnarray}
\label{1}
\pa  & =& \frac{\pa}{\pa z}
\quad , \quad
D \ = \ \frac{\pa}{\pa \th}  +  \frac{1}{2} \, \tb \pa
\qquad \ , \quad  \
\dab \ = \ \frac{\pa}{\pa \tb}  +  \frac{1}{2} \, \th  \pa
\\
\pab & =& \frac{\pa}{\pa \zb}
\quad , \quad
D_- \ = \ \frac{\pa}{\pa \thm}  +  \frac{1}{2} \, \tbm \pab
\ \  , \quad
\dab_- \ = \ \frac{\pa}{\pa \tbm}  +  \frac{1}{2} \, \thm  \pab
\ \ .
\nonumber
\end{eqnarray}
The graded Lie brackets between these vector fields are given by
\begin{equation}
\label{2}
\{  D ,  \dab \} =  \pa
\qquad , \qquad
\{  D_-  ,  \dab_- \} =  \pab
\ \ \ ,
\end{equation}
all others brackets being zero, in particular,
\begin{equation}
\label{3}
D^2   =  0  = \dab^2
\qquad , \qquad
( D_- ) ^2   =  0  =
( \dab_- ) ^2
\ \ .
\end{equation}
For later reference, we note that this set of equations implies
\begin{equation}
[D,\dab ]^2 = \pa^2
\qquad , \qquad
[D_-,\dab_- ]^2 = \pab^2
\ \ .
\end{equation}

The cotangent vectors which are dual to the canonical
tangent vectors (\ref{1})
are given by the 1-forms
\begin{eqnarray}
\label{9}
e^z & =& dz  +  \frac{1}{2} \, \th  d\tb  +
\frac{1}{2} \, \tb  d\th
\qquad \quad , \quad \
e^{\th} = d \th
\quad \ \ \  , \quad  \
e^{\tb}= d \tb
\\
e^{\zb} & = &  d \zb
 +  \frac{1}{2} \, \thm  d\tbm  +
\frac{1}{2} \, \tbm d\thm
\ \  , \quad
e^{\thm} = d \thm
\quad , \quad
e^{\tbm}= d \tbm
\nonumber
\end{eqnarray}
and that the graded commutation relations (\ref{2})(\ref{3}) are
equivalent to the {\em structure equations}
\begin{eqnarray}
\label{10}
0 & = & de^z  +  e^{\th} \, e^{\tb}
\quad \quad , \quad  \quad  \
de^{\th}   = 0 =
de^{\tb}
\\
0 & = & de^{\zb}  +  e^{\thm} \, e^{\tbm}
\quad , \qquad
de^{\thm}   = 0 =
de^{\tbm}
\ \ .
\nonumber
\end{eqnarray}

\subsubsection{Superconformal transformations}

By definition of the SRS,
any two sets of local coordinates, say $( {\zz} ; \bar{\zz} )$
and $( {\zz}^{\prime} ; \bar{\zz}^{\prime} )$,
are related by a
superconformal transformation, i.e.
a mapping for which
$D, \, \dab$ transform
among themselves and similarly
$D_-, \, \dab_-$:
\begin{eqnarray}
\label{3f}
D & = & [\,
D \th^{\prime} \, ] \, D ^{\prime} \, + \,
[ \, D \tb^{\prime} \, ] \, \dab ^{\prime}
\quad , \quad
D_- \ = \ [\,
D_- \th^{-\prime} \, ] \, D_- ^{\prime} \, + \,
[ \, D_- \tb^{-\prime} \, ] \, \dab_- ^{\prime}
\\
\dab & = & [\,
\dab \th^{\prime} \, ] \, D ^{\prime} \, + \,
[ \, \dab \tb^{\prime} \, ] \, \dab ^{\prime}
\quad , \quad
\dab_- \ = \ [\,
\dab_- \th^{-\prime} \, ] \, D_- ^{\prime} \, + \,
[ \, \dab_- \tb^{-\prime} \, ] \, \dab_- ^{\prime}
\ \ .
\nonumber
\end{eqnarray}
These properties are equivalent to the following two conditions :

\noindent (i)
\begin{eqnarray}
\label{4f}
{\zz}^{\prime} & = & {\zz} ^{\prime} ( {\zz} )
\quad \Longleftrightarrow \quad
D_- {\zz}^{\prime}  = 0 =
\dab_- {\zz}^{\prime}
\\
\bar{\zz} ^{\prime} &  = &  \bar{\zz}  ^{\prime} (\bar{\zz} )
\quad \Longleftrightarrow \quad
D \bar{\zz} ^{\prime} = 0 =
\dab \bar{\zz} ^{\prime}
\ \ ,
\nonumber
\end{eqnarray}

\noindent (ii)
\begin{eqnarray}
\label{5f}
D z^{\prime} & = & \frac{1}{2}
\th ^{\prime} (D \tb ^{\prime} ) +  \frac{1}{2}
\tb ^{\prime}  (D \th ^{\prime} )
\qquad \qquad , \quad
\dab z^{\prime} \ = \ \frac{1}{2}
\th ^{\prime}  (\dab \tb ^{\prime} ) +  \frac{1}{2}
\tb ^{\prime}  (\dab \th ^{\prime} )
\\
D_- \zb^{\prime} & = & \frac{1}{2}
\th ^{-\prime} (D_- \tb ^{-\prime} ) +  \frac{1}{2}
\tb ^{-\prime}  (D_- \th ^{-\prime} )
\ \ , \ \
\dab_- \zb^{\prime} \ = \ \frac{1}{2}
\th ^{-\prime}  (\dab_- \tb ^{-\prime} ) +  \frac{1}{2}
\tb ^{-\prime}  (\dab_- \th ^{-\prime} )
.
\nonumber
\end{eqnarray}
Application of the algebra (\ref{2})(\ref{3}) to eqs.(\ref{5f})
yields a set of integrability conditions,
\begin{eqnarray}
0 & = &
(D \th^{\prime} \, ) \,
( D \tb^{\prime} \, )
\nonumber
\\
0 & = &
(\dab \tb^{\prime} \, ) \,
( \dab \th^{\prime} \, )
\label{3h}
\\
0 & = &
(D \th^{\prime}  ) \,
( \dab\tb^{\prime}  )  +
(D \tb^{\prime}  ) \,
( \dab \th^{\prime}  ) \, - \,
\left[ \, \pa z^{\prime}
 +  \frac{1}{2} \, \tb^{\prime} \, \pa \th^{\prime}
 +  \frac{1}{2} \, \th^{\prime} \, \pa \tb^{\prime}
\, \right]
\nonumber
\end{eqnarray}
(and similarly for the $\zb$-sector).
Obviously, there are four possibilities to satisfy the first two
of these equations. The two solutions $D\th^{\prime} = 0 =
\dab \th^{\prime}$ and
$\dab \tb^{\prime} = 0 =
D\tb^{\prime}$
are not acceptable, because they would
imply that the change of coordinates
is non-invertible (the associated Berezinian would vanish).
The third possibility,
$D\th^{\prime} = 0 =
\dab \tb^{\prime}$
amounts to interchanging the r\^ole of $\th$ and $\tb$, since
it leads to
$D \propto \dab ^{\prime}$ and
$\dab \propto D^{\prime}$.
The remaining solution is
\begin{equation}
\label{3i}
D \tb^{\prime} \ = \ 0 \ =
\dab \th^{\prime}
\ \ \ ,
\end{equation}
which implies that $D$ and $\dab$ separately transform
into themselves. The resulting transformation laws can be written as
\begin{eqnarray}
D ^{\prime} & = & {\rm e} ^w \ D
\nonumber
\\
\dab ^{\prime} & = & {\rm e} ^{\bar{w}} \ \dab
\label{7a}
\\
\pa ^{\prime} & = &
\{  D^{\prime}  ,  \dab ^{\prime}  \}  =
{\rm e} ^{w+\bar{w}}  \, [  \pa   +
( \dab w )  D  +
(D \bar{w} )  \dab   ]
\nonumber
\end{eqnarray}
with
\begin{eqnarray}
\label{8a}
{\rm e}^{-w} & \equiv & D \th^{\prime}
\ \ \ \ \ , \ \ \ \ \
D w \ = \ 0
\\
{\rm e}^{-\bar{w}} & \equiv & \dab  \tb^{\prime}
\ \ \ \ \ , \ \ \ \ \
\dab  \bar{w} \ = \ 0
\ \ \ .
\nonumber
\end{eqnarray}
The last equation in (\ref{3h}) then leads to
\begin{equation}
\label{w}
{\rm e}^{-w-\bar{w}} \ = \
\pa z^{\prime}
\, + \, \frac{1}{2} \, \tb^{\prime} \, \pa \th^{\prime}
\, + \, \frac{1}{2} \, \th^{\prime} \, \pa \tb^{\prime}
\ \ \ .
\end{equation}
In the remainder of the text, {\em superconformal transformations}
are assumed to satisfy conditions (\ref{4f})(\ref{5f}) and
(\ref{3i}). Analogous equations hold in the $\zb$-sector,
\begin{eqnarray}
\label{8b}
D_- ^{\prime} &=& {\rm e}^{w^-} D_-
\qquad , \qquad
{\rm e}^{-w^-} \equiv D_-\th^{-\prime}
\qquad , \qquad
D_-w^- =0
\\
\dab_- ^{\prime} &=& {\rm e}^{\bar{w} ^-} \dab_-
\qquad , \qquad
{\rm e}^{-\bar{w} ^-} \equiv \dab_- \tb^{-\prime}
\qquad , \qquad
\dab_- \bar{w} ^-  =0
\nonumber
\end{eqnarray}
with the relation
\begin{equation}
\label{ww}
{\rm e}^{-w^--\bar{w} ^-} = \pab \zb^{\prime}
+{1\over 2} \tb^{-\prime}  \pab \th^{-\prime}
+{1\over 2} \th^{-\prime} \pab \tb^{-\prime}
\ \ .
\end{equation}

To conclude our discussion,
we note that
the superconformal transformations
of the canonical 1-forms read
\begin{eqnarray}
\label{12a}
e^{z^{\prime}} & = & {\rm e} ^{-w-\bar{w}} \, e^z
\qquad \qquad \qquad , \qquad
e^{\zb^{\prime}} \ \ = \ {\rm e} ^{-w^--\bar{w}^-} \, e^{\zb}
\\
e^{\th ^{\prime}} & = & {\rm e} ^{-w}  \, [  e^{\th}  -  e^z
( \dab w ) ]
\qquad , \qquad
e^{\th ^{-\prime}} \ = \ {\rm e} ^{-w^-}  \, [  e^{\thm}  -  e^{\zb}
( \dab_- w^- ) ]
\nonumber
\\
e^{\tb^{\prime}} & = & {\rm e} ^{-\bar{w}}  \, [  e^{\tb}  -  e^{z}
(D \bar{w} ) ]
\qquad , \qquad
e^{\tb^{-\prime}} \ = \ {\rm e} ^{-\bar{w}^-}  \, [e^{\tbm} - e^{\zb}
(D_- \bar{w}^- ) ]
\nonumber
\end{eqnarray}
with
$w, \bar{w}$ and $w^-, \bar{w}^-$ given by eqs.(\ref{8a}) and (\ref{8b}),
respectively.

\subsubsection{$U(1)$-symmetry and
complex conjugation}

The $N=2$ supersymmetry algebra
admits a $U(1) \otimes U(1)$ automorphism group.
In the {\em Minkowskian framework},
the latter may be viewed
as $SO(1,1) \otimes SO(1,1)$ in which case the Grassmannian coordinates
$\th, \tb, \th^- , \tb^-$ are all real and independent
or it may be regarded as $SO(2) \otimes SO(2)$ in which case the
Grassmannian coordinates are complex and related by
$\th^{\ast} = \tb$ and
$(\th^-)^{\ast} = \tb^-$.

\section{Projection to component fields}

A generic $N=2$ superfield admits the $\th$-expansion
\begin{eqnarray}
\nonumber
F({\zz}\, ; \bar{{\cal Z}} ) &=&
 a+ \th \alpha + \tb \beta + \thm \gamma + \tbm \delta \\
\nonumber &&+
\th\tb b+ \th\thm c+\th\tbm d+\tb\thm e+\tb\tbm f+\thm\tbm g \\
\nonumber &&+
\th \tb
\thm \epsilon +\th\tb\tbm \zeta +\th\thm\tbm \eta +\tb\thm\tbm \la \\
 &&+
\th\tb\thm\tbm h
\ \ ,
\end{eqnarray}
where the component fields $a,\alpha, \beta,...$ depend on
$z$ and $\zb$. Equivalently, these space-time fields can be introduced
by means of projection,
\begin{eqnarray}
F \! \mid &=& a
\nonumber
\\
DF \! \mid &=& \alpha \ \ \ \ , \ \ \ \ \dab F \! \mid =\beta \ \
\ \ , \ \ \ \  D_- F \! \mid \, = \gamma
\ \ \ \ , \ \ \ \  \dab_- F \! \mid \, = \delta
\nonumber
\\
{[ D,\dab ]} F \! \mid &=& -2b  \ \ \ \ \ \ ,  \ \ \ \ \ \
DD_- F \! \mid \, =-c \ \ \ \ \ \  , \ \ \ \ \ \ \ D\dab_- F \!
\mid \, = -d
\nonumber
\\
\dab D_-F\! \mid &=&
-e  \ \ \ \ \ \ \ \, , \ \ \ \ \ \ \dab \dab_-F\!
 \mid \, =-f
\ \ \ \ \ ,\ \ \ \ \ \ \ [ D_-,\dab_-] F\! \mid \, = -2g
\nonumber \\
{[ D,\dab ]} D_- F \! \mid &=& -2 \epsilon
\ \ \ \ \ \  ,  \ \ \ \ \ \
{[ D,\dab ]} \dab_- F\! \mid \, = -2 \zeta
\nonumber \\
D [ D_-,\dab_-] F \! \mid &=& -2 \eta  \ \ \ \ \ \; , \ \ \ \
\dab [ D_-,\dab_-] F\! \mid \, = -2 \la
\\
{[ D,\dab ] [ D_-,\dab_-]} F\! \mid &=& 4h \ \ \ ,
\nonumber
\end{eqnarray}
where the bar denotes the projection onto the lowest component
of the corres\-ponding superfield.

\chapter{(2,0) Theory}

In this chapter, we discuss
(2,0) SRS's
and super Beltrami differentials.
The projection of superspace results
to ordinary space will be performed in the end.

\section{(2,0) Super Riemann Surfaces}

A $(2,0)$ SRS
is locally parametrized by coordinates
$(z, \zb , \th , \tb )$, the notation being the same as the one
for the $N=2$ theory discussed in the last chapter.
The basic geometric quantities and
relations are obtained from those of the $N=2$ theory by dropping the
terms involving $\thm$ and $\tbm$. Thus, in the
$z$-sector, one has the same equations as in the $N=2$ case.
For later reference, we now summarize all relations which hold
in the present case in terms of a generic system of coordinates
$(Z, \ZB , \TH , \TB )$.

The canonical basis of the tangent space and of the cotangent space
are respectively given by
\begin{equation}
\pa_Z = \frac{\pa}{\pa Z}
\quad , \quad
\pa_{\ZB} = \frac{\pa}{\pa \ZB}
\quad , \quad
D_{\TH} = \frac{\pa}{\pa \TH} + {1\over 2} \, \TB \pa_Z
\quad , \quad
D_{\TB}  = \frac{\pa}{\pa \TB} + {1\over 2} \, \TH \pa_Z
\end{equation}
and
\begin{equation}
e^Z = dZ + {1\over 2} \, \TH d\TB + {1\over 2} \, \TB d\TH
\quad , \quad
e^{\ZB} = d\ZB
\quad , \quad
e^{\TH} = d\TH
\quad , \quad
e^{\TB} = d\TB
\ \ ,
\label{cota}
\end{equation}
the {\em structure relations} having the form
\begin{equation}
\{D_{\TH} , D_{\TB} \} = \pa_Z
\qquad , \qquad
(D_{\TH}) ^2 \ = 0 = ( D_{\TB}) ^2
\qquad , \quad
...
\end{equation}
and
\begin{equation}
\label{strr}
0 = de^Z + e^{\TH} e^{\TB}
\qquad , \qquad
0 = de^{\ZB} = de^{\TH} = de^{\TB}
\ \ .
\end{equation}
A change of coordinates
$(Z, \ZB , \TH, \TB) \to
(Z^{\prime} , \ZB^{\prime}  , \TH^{\prime} , \TB^{\prime})$
is a {\em superconformal transformation} if it
satisfies the conditions
\begin{eqnarray}
Z^{\prime} & = & Z^{\prime}(Z , \TH , \TB)
\quad \Longleftrightarrow \quad
0=
\pa_{\ZB} Z^{\prime}
\nonumber
\\
\TH^{\prime} & =& \TH^{\prime}(Z , \TH , \TB)
\quad \Longleftrightarrow \quad
0= \pa_{\ZB} \TH^{\prime}
\label{4}
\\
\TB^{\prime} & = & \TB^{\prime}(Z , \TH , \TB)
\quad \Longleftrightarrow \quad
0= \pa_{\ZB} \TB^{\prime}
\nonumber
\\
\ZB^{\prime} &  = &  \ZB ^{\prime} (\ZB )
\quad \quad \ \,
\quad \Longleftrightarrow \quad
0=
D_{\TH} \ZB^{\prime}  =
D_{\TB} \ZB^{\prime}
\nonumber
\end{eqnarray}
and
\begin{eqnarray}
\label{5}
D_{\TH} Z^{\prime} & = & \frac{1}{2} \,
\TH ^{\prime}  (D_{\TH} \TB ^{\prime} ) +  \frac{1}{2} \,
\TB ^{\prime}  (D_{\TH} \TH ^{\prime} )
\\
D_{\TB} Z^{\prime} & = & \frac{1}{2} \,
\TH ^{\prime}  (D_{\TB} \TB ^{\prime} ) +  \frac{1}{2} \,
\TB ^{\prime}  (D_{\TB} \TH ^{\prime} )
\ \ ,
\nonumber
\end{eqnarray}
as well as
\begin{equation}
\label{3c}
D_{\TH} \TB^{\prime} \ = \ 0 \ =
D_{\TB} \TH^{\prime}
\ \ .
\end{equation}
The induced change of the canonical tangent and cotangent vectors reads
\begin{eqnarray}
D_{\TH} ^{\prime} & = & {\rm e} ^W \, D_{\TH}
\qquad , \qquad
\pa_{Z} ^{\prime} \ = \
{\rm e} ^{W+\bar{W}}  \, [  \pa _Z  +
(D_{\TB} W )  D_{\TH}  +
(D_{\TH} \bar{W} )  D_{\TB}  ]
\nonumber
\\
D_{\TB} ^{\prime} & = & {\rm e} ^{\bar{W}} \, D_{\TB}
\qquad , \qquad
\pa_{\ZB}^{\prime}  \ = \  (\pa_{\ZB} \ZB^{\prime} )^{-1} \,
\pa_{\ZB}
\label{12}
\end{eqnarray}
and
\begin{eqnarray}
e^{Z^{\prime}} & = & {\rm e} ^{-W-\bar{W}} \, e^Z
\qquad \, , \qquad
e^{\TH ^{\prime}} \ = \ {\rm e} ^{-W}  \, [  e^{\TH}  -  e^Z \,
(D_{\TB} W ) ]
\nonumber
\\
e^{\ZB^{\prime}} & = &
(\pa_{\ZB}  \ZB^{\prime} ) \, e^{\ZB}
\qquad \ , \qquad\,
e^{\TB ^{\prime}} \ = \ {\rm e} ^{-\bar{W}}  \, [  e^{\TB}  -  e^Z \,
(D_{\TH} \bar{W} ) ]
\label{12m}
\end{eqnarray}
with
\begin{eqnarray}
\label{8}
{\rm e}^{-W} & \equiv & D_{\TH} \TH^{\prime}
\ \ \ \ \ , \ \ \ \ \
D_{\TH} W \, = \, 0
\\
{\rm e}^{-\bar{W}} & \equiv & D_{\TB} \TB^{\prime}
\ \ \ \ \ , \ \ \ \ \
D_{\TB} \bar{W} \, = \, 0
\nonumber
\end{eqnarray}
and
\begin{equation}
{\rm e}^{-W-\bar{W}}  =
\pa_Z Z^{\prime}
 +  \frac{1}{2} \, \TB^{\prime}  \pa_Z \TH^{\prime}
 +  \frac{1}{2} \, \TH^{\prime}  \pa_Z \TB^{\prime}
\ \ .
\end{equation}

In the Euclidean framework, $\TH$ and $\TB$ are independent
complex variables and the action functional will also represent a
complex quantity. In the Minkowskian setting,
one either deals with real independent coordinates $\TH$ and $\TB$
($SO(1,1)$ automorphism group) or with complex conjugate
variables $\TH$ and $\TH^{\ast} = \TB$ ($SO(2)$ automorphism group).

\section{Beltrami superfields and U(1)-symmetry}

Beltrami (super)fields parametrize (super)conformal structures
with respect to a given (super)conformal structure.
Thus, we start from a reference complex structure corresponding
to a certain choice of local coordinates
$(z , \zb , \th , \tb )$ for which we denote the canonical tangent
vectors by
\[
\pa  = \frac{\pa}{\pa z} \ \ \  ,  \ \ \
\pab  = \frac{\pa}{\pa \zb} \ \ \  ,  \ \ \
D
\equiv  D_{\th}  = \frac{\pa}{\pa \th}  +  \frac{1}{2} \, \tb  \pa
\ \ \  ,  \ \ \
\dab  \equiv  D_{\tb}  = \frac{\pa}{ \pa \tb}  +  \frac{1}{2} \, \th
\pa
\ \ .
\]
Then, we pass over to an arbitrary complex structure (corresponding
to local coordinates
$(Z , \ZB , \TH , \TB )$) by a smooth change of coordinates
\begin{equation}
\label{13}
(z ,  \zb  ,  \th  , \bar{\th})
\longrightarrow
\left(  Z
(z ,  \zb  ,  \th  , \bar{\th})  ,
\ZB
(z ,  \zb  ,  \th  , \bar{\th})  ,
\TH
(z ,  \zb  ,  \th  , \bar{\th})  ,
\TB
(z ,  \zb  ,  \th  , \bar{\th})  \right)
\ \ .
\end{equation}
To simplify the notation, we label the small coordinates
by small indices $a,\, b $, e.g.
$(e^a ) = ( e^z , e^{\zb} , e^{\th} , e^{\tb} ) ,\
(D_a ) = ( \pa , \pab , D , \dab )$
and the capital
coordinates by capital indices $A,\, B$.

The transformation of the canonical 1-forms induced by the change
of coordinates (\ref{13}) reads
\[
e^B \ = \   \sum_{a=z,\zb ,\th , \tb}  e^a \, E_a ^{\ B}
\ \ \ \ \ \ \ {\rm for} \ \ \ B \, = \, Z , \ZB , \TH , \TB
\ \ \ .
\]
Here, the $E_a ^{\ B}$ are superfields whose explicit form is easy to
determine from the expressions (\ref{cota}) and $d=e^a D_a$:
for $a=z,\zb,\th,\tb$, one finds
\begin{eqnarray}
E_a ^{\ Z} & = & D_a Z \, - \, \frac{1}{2} \, (D_a \TH) \TB
\, - \, \frac{1}{2} \, (D_a \TB) \TH
\label{13a}
\\
E_a ^{\ \TH} & = & D_a \TH \quad , \quad
E_a ^{\ \TB} \ = \ D_a \TB\quad , \quad
E_a ^{\ \ZB} \ = \ D_a \ZB
\ \ .
\nonumber
\end{eqnarray}
Since $e^Z$ and $e^{\ZB}$ transform homogeneously under the
superconformal transformations (\ref{4})-(\ref{3c}), one can extract
from them some Beltrami variables
$H_a^{\ b}$ which are inert under these transformations: to do so, we
factorize $E_z ^{\ Z}$ and $E_{\zb} ^{\ \ZB}$ in
$e^Z$ and $e^{\ZB}$, respectively :
\begin{equation}
\label{14}
e^Z  =    [ \, e^z \, + \sum_{a\neq z}
e^{a} \, H_{a} ^{\ z}  \, ]
\, E_z ^{\ Z}
\quad , \quad
e^{\ZB}  =    [ \, e^{\zb} \,  + \sum_{a\neq \zb}
e^{a} \, H_{a} ^{\ \zb}  \, ]
\, E_{\zb} ^{\ \ZB}
\end{equation}
with
\begin{equation}
\label{15}
H_a ^{\ z}  \equiv  \frac{E_a ^{\ Z}}{E_z ^{\ Z}}  \ \ \ \
{\rm for} \ a  =  \zb  ,  \th  ,  \tb
\quad {\rm and} \quad
H_a ^{\ \zb}  \equiv  \frac{E_a ^{\ \ZB}}{E_{\zb} ^{\ \ZB}}  \ \ \ \
{\rm for} \ a  =  z  ,  \th  ,  \tb
\ .
\end{equation}
By construction,
$E_a ^{\ Z}$ and $E_a ^{\ \ZB}$ vary homogeneously under the
transformations (\ref{4})-(\ref{3c}), in particular
\[
E_z ^{\ Z^{\prime}} \ = \ {\rm e}^{-W-\bar{W}} \ E_z ^{\ Z}
\ \ \ .
\]
This transformation law
and the index structure of $E_z^{\ Z}$ advises us to decompose
this complex variable as
\begin{equation}
\label{16}
E_z ^{\ Z} \ \equiv \ \LA_{\th} ^{\ \TH} \; \LB_{\tb} ^{\ \TB}
\ \equiv \ \LA \; \LB
\end{equation}
with $\LA , \LB$ transforming according to
\begin{equation}
\label{17}
\LA ^{\TH^{\prime}}\ = \ {\rm e}^{-W} \
\LA ^{\TH}
\ \ \ \ \ , \ \ \ \ \
\LB ^{\TB^{\prime}}\ = \ {\rm e}^{-\bar{W}} \
\LB ^{\TB}
\ \ \ .
\end{equation}
Then, we can use $\LA$ and $\LB$ to extract Beltrami coefficients
from
$e^{\TH}$ and $e^{\TB}$, respectively, in analogy to $N=1$ supersymmetry
\cite{dg} :
\begin{equation}
H_a ^{\ \th} \ = \ \frac{1}{\LA} \ [ \, E_a ^{\ \TH} \; - \;
H_a ^{\ z} \, E_z ^{\ \TH} \, ]
\ \ \ \  ,  \ \ \ \
H_a ^{\ \tb} \ = \ \frac{1}{\LB} \ [ \, E_a ^{\ \TB} \; - \;
H_a ^{\ z} \, E_z ^{\ \TB} \, ]
\ \ \ \ \ {\rm for} \ \ a \, = \, \zb , \th , \tb
\ \ .
\label{ana}
\end{equation}
The {\em final result} is best summarized in matrix form,
\begin{equation}
\label{18}
\left( \ e^Z \ ,\ e^{\bar{Z}} \ ,\ e^{\TH} \ , \ e^{\TB} \ \right) \ =\
\left( \ e^z \ ,\ e^{\bar{z}} \ ,\ e^{\th} \ ,\ e^{\tb} \ \right) \ \cdot
M \cdot Q
\end{equation}
with
\begin{equation}
\label{19}
M \ = \ \left( \begin{array}{clccr}
1           & {H_z}^{\zb}     & 0       & 0                \\
{H_{\zb}}^z & 1               & {H_{\zb}}^{\th}  &  H_{\zb} ^{\ \tb} \\
\HT& {H_{\th}}^{\zb}&{H_{\th}}^{\th}& {H_{\th}}^{\tb}
\\
\HB& {H_{\tb}}^{\zb}&{H_{\tb}}^{\th}& {H_{\tb}}^{\tb}
\end{array}   \right)
\ \ \ \ \ , \ \ \ \ \
Q      \ = \ \left( \begin{array}{clccr}
\LA \LB & 0     &   \tau   &  \bt \\
0     & \OM   & 0     & 0        \\
0     & 0     & \LA  &  0   \\
0     & 0     & 0    & \LB
\end{array}   \right)
\end{equation}
where
\begin{equation}
\label{20}
\OM \equiv \OM_{\zb} ^{\ \ZB}
\equiv E_{\zb} ^{\ \ZB}
\ \ \ \ , \ \ \ \
\tau \equiv
\tau_{z} ^{\ \TH} \equiv
E_{z} ^{\ \TH}
\ \ \ \ , \ \ \ \
\bt \equiv
\bt_{z} ^{\ \TB} \equiv
E_{z} ^{\ \TB}
\ \ \ .
\end{equation}
All the `$H$' are invariant under the superconformal transformations
(\ref{4})-(\ref{3c}). Under the latter, the factors $\LA, \LB$ change
according to
eqs.(\ref{17}) while
$\Omega$ and $\tau , \bt$ vary according to
$\Omega^{\ZB^{\prime}} = \Omega^{\ZB} \pa \ZB^{\prime} / \pa \ZB$ and
\begin{eqnarray}
\label{21}
\tau^{\TH ^{\prime}} & = & {\rm e} ^{-W}  \ [ \, \tau ^{\TH} \ - \
\LA ^{\TH} \, \LB^{\TB} \ (D_{\TB} W ) \,]
\\
\bt^{\TB ^{\prime}} & = & {\rm e} ^{-\bar{W}} \ [ \, \bt ^{\TB} \ - \
\LA ^{\TH} \, \LB^{\TB}  \  (D_{\TH} \bar{W} ) \,]
\ \ \ .
\nonumber
\end{eqnarray}

Obviously, the decomposition (\ref{16}) has introduced a U(1)-symmetry
which leaves $e^Z , e^{\ZB} , e^{\TH} , e^{\TB}$ invariant and
which is given by
\begin{eqnarray}
\label{22}
\LA ^{\prime} & = & {\rm e} ^K \ \LA
\ \ \ \ \ \ \ \ \ , \ \ \ \ \ \ \ \ \ \ \ \
\LB ^{\prime} \ = \ {\rm e} ^{-K} \ \LB
\\
(H_a ^{\ \tb} )^{\prime} & = &
{\rm e}^{K} \
H_a ^{\ \tb}
 \ \ \ \ \ \ ,  \ \ \ \ \ \ \
(H_a ^{\ \th} )^{\prime} \ = \
{\rm e}^{-K} \
H_a ^{\ \th}
\ \ \ \ \ {\rm for} \ \ a \, = \, \zb , \th , \tb
\ \ ,
\nonumber
\end{eqnarray}
where $K$ is an unconstrained superfield. In the sequel, we will
encounter this symmetry in other places and forms.

Besides the transformations we have considered so far, there are the
superconformal variations of the small coordinates under which the
basis 1-forms change according to
\begin{eqnarray}
\label{23}
e^{z^{\prime}} & = & {\rm e} ^{-w-\bar{w}} \ e^z
\qquad , \qquad
e^{\th ^{\prime}} \ = \ {\rm e} ^{-w}  \ [ \, e^{\th} \ - \ e^z \
(\dab w ) \,]
\\
e^{\zb^{\prime}} & = &
e^{\zb} \ \pab \zb^{\prime}
\ \qquad \quad , \quad\quad
e^{\tb ^{\prime}} \ =\  {\rm e} ^{-\bar{w}}  \ [ \, e^{\tb} \ - \ e^z \
(D \bar{w} ) \,]
\nonumber
\end{eqnarray}
with $D w \, = \, 0 \, = \, \dab \bar{w}$.
The determination of the induced transformations of the `$H$' and of
$\LA, \LB , \OM , \tau , \bt$ is straightforward
and we only present the results to which we will refer later on.
In terms of the quantity
\[
Y =  1 + (\dab w) \, \HT + (D \bar{w}) \HB
\ \ ,
\]
the combined
superconformal and $U(1)$ transformation laws have the form
\begin{eqnarray}
\LA ^{\prime} & = & {\rm e} ^{K} \,
{\rm e} ^{w} \, Y^{1/2} \, \LA
\qquad , \qquad
\bar{\LA} ^{\prime} \ = \ {\rm e} ^{-K} \,
{\rm e} ^{\bar w} \, Y^{1/2} \, \bar{\LA}
\qquad , \qquad
\Omega ^{\prime} \ = \ (\pab \zb ^{\prime} )^{-1}  \, \Omega
\nonumber  \\
H_{\th^{\prime}} ^{\ \, z^{\prime}} & = & {\rm e}^{- \bar w} \,
Y^{-1} \, \HT
\qquad , \qquad
H_{\tb^{\prime}} ^{\ \, z^{\prime}} \ = \ {\rm e}^{-w} \,
Y^{-1} \, \HB
\nonumber
\\
H_{\th^{\prime}} ^{\ \, \tb^{\prime}} & = & {\rm e}^{+K} \,
{\rm e}^{+w- \bar w} \,
Y^{-1/2} \, \left\{ H_{\th} ^{\ \tb}  + Y^{-1}
[ \, (\dab w)\, H_{\th}^{\ \tb} + (D\bar w ) H_{\tb}^{\ \tb}
] \HT \right\}
\nonumber  \\
H_{\tb^{\prime}} ^{\ \, \th^{\prime}} & = & {\rm e}^{-K} \,
{\rm e}^{-w + \bar w} \,
Y^{-1/2} \, \left\{ H_{\tb} ^{\ \th}  + Y^{-1}
[ \, (D \bar w)\, H_{\tb}^{\ \th} + (\dab w ) H_{\th}^{\ \th}
] \HB \right\}
\nonumber  \\
H_{\tb^{\prime}} ^{\ \, \tb^{\prime}} & = &
{\rm e}^{+K} \,
Y^{-1/2} \, \left\{ H_{\tb} ^{\ \tb}  + Y^{-1}
[ \, (D \bar w)\, H_{\tb}^{\ \tb} + (\dab  w ) H_{\th}^{\ \tb}
] \HB \right\}
\nonumber  \\
H_{\th^{\prime}} ^{\ \, \th^{\prime}} & = & {\rm e}^{-K} \,
Y^{-1/2} \, \left\{ H_{\th}^{\ \th}  + Y^{-1}
[ \, (\dab w)\, H_{\th}^{\ \th} + (D\bar w ) H_{\tb}^{\ \th}
] \HT \right\}
\nonumber  \\
H_{\zb^{\prime}} ^{\ \, z^{\prime}} & = & {\rm e}^{-w-\bar{w}}
\, (\pab \zb ^{\prime} )^{-1}  \, Y^{-1} \, \HZB
\label{24}
\\
H_{\th^{\prime}} ^{\ \, \zb^{\prime}} & = & {\rm e}^{w}
\, (\pab \zb ^{\prime} ) \, H_{\th}^{\ \zb}
\qquad , \qquad
H_{\tb^{\prime}} ^{\ \, \zb^{\prime}} \ = \ {\rm e}^{\bar w}
\, (\pab \zb ^{\prime} ) \, H_{\tb}^{\ \zb}
\nonumber  \\
H_{z^{\prime}} ^{\ \, \zb^{\prime}} & = & {\rm e}^{w + \bar{w}}
\, (\pab \zb ^{\prime} ) \left[ H_z^{\ \zb} +
(\dab w) H_{\th} ^{\ \zb} +
(D \bar w) H_{\tb} ^{\ \zb} \right]
\ \ .
\nonumber
\end{eqnarray}
The given variations of $\LA, \LB$
and $H_a^{\ \th} , H_a ^{\ \tb}$ result from a symmetric splitting
of the transformation law
\[
(\LA \LB )^{\prime} = {\rm e}^{w+ \bar w} Y (\LA \LB )
\ \ .
\]
The ambiguity involved in this decomposition is precisely
the $U(1)$-symmetry (\ref{22}):
\[
\LA ^{\prime} = {\rm e}^K {\rm e}^w Y^{1/2} \LA
\quad , \quad
\LB ^{\prime} = {\rm e}^{-K} {\rm e}^{\bar w} Y^{1/2} \LB
\ \ .
\]

Due to the structure relations (\ref{strr}), not all of the
{\em super Beltrami coefficients} $H_a ^{\ b}$ and of the
{\em integrating factors}
$\LA, \LB , \OM , \tau , \bt$ are independent variables.
For instance, the structure relation
$0 = d e^{\ZB}$ is equivalent to the set of equations
\begin{eqnarray}
0 & = & ( \, D_a \, - \, H_{a} ^{\ \zb} \, \pab  \, - \, \pab
H_a ^{\ \zb} \, ) \, \OM
\ \ \ \ \ \ \ \ \ \ \ \ {\rm for } \ \ \ a = z   , \th , \tb
\nonumber
\\
0 & = & D_a (H_z ^{\ \zb}  \OM ) \ - \ \pa ( H_a ^{\ \zb } \OM  )
\ \ \ \ \
\ \ \ \ \ \ \ \ \ \ {\rm for } \ \ \ a =  \th , \tb
\nonumber
\\
0 & = & D (H_{\th} ^{\ \zb}  \OM )
\nonumber
\\
0 & = & \dab (H_{\tb} ^{\ \zb}  \OM )
\label{app}
\\
0 & = & \dab (H_{\th} ^{\ \zb}  \OM )  \ + \
D (H_{\tb} ^{\ \zb}  \OM )  \ - \ H_z ^{ \ \zb} \OM
\ \ .
\nonumber
\end{eqnarray}
The last equation can be solved for $H_z ^{\ \zb}$ and the two
equations
preceding it provide constraints for the fields $H_{\th} ^{\ \zb},
\, H_{\tb} ^{\ \zb}$.

In summary, by solving all resulting
equations
which are algebraic, we
find the following result. In the
$\zb$-sector, there is
one integrating factor
($\OM$) and two independent
Beltrami superfields ($H_{\th} ^{\ \zb}$ and $H_{\tb} ^{\ \zb} $),
each of which satisfies a constraint reducing the number of its
independent component fields by a factor 1/2.
In section 3.9,
the constraints on $H_{\th}^{\ \zb}$ and $H_{\tb}^{\ \zb}$
will be explicitly solved in terms of `prepotential' superfields
$H^{\zb}$ and $\hat{H} ^{\zb}$.
In the
$z$-sector, there are two
integrating factors
($\LA, \, \LB$) and four independent and unconstrained Beltrami variables
($H_{\zb} ^{\ z}, \, \HT , \, \HB$ and a non-U(1)-invariant
combination of
$H_{\th} ^{\ \th} , \,
H_{\tb} ^{\ \tb}$, e.g.
$H_{\th} ^{\ \th} /
H_{\tb} ^{\ \tb}$).
The dependent Beltrami fields only depend on the others
and {\em not} on the integrating factors.
This is an important point, since
the integrating factors represent non-local
functionals of the `$H$' by virtue of the  differential
equations that they satisfy, see below.

To be more explicit,
in the $z$-sector, one finds
\begin{eqnarray}
H_{\tb} ^{\ \th} H_{\tb} ^{\ \tb}
& = & -\,
(\dab  -  \HB  \pa )
\HB
\ \ \ \ \ , \ \ \ \ \
H_{\th} ^{\ \tb} \HO
\ = \ -\,  (D  -  \HT  \pa )
 \HT
\nonumber
\\
\HO  H_{\tb} ^{\ \tb} \, + \, H_{\tb} ^{\ \th}  H_{\th} ^{\ \tb}
& = & 1\, - \,
(  \dab  -  \HB  \pa  )   \HT  \, - \,
(  D  -  \HT  \pa  )   \HB
\nonumber
\\
\HZ H_{\th} ^{\ \tb} +
H_{\zb}^{\ \tb} \HO
 & = &
(  D  -  \HT  \pa  ) \HZB \, - \,
(\pab - \HZB \pa ) \HT
\label{26}
\\
\HZ H_{\tb} ^{\ \tb} +
H_{\zb}^{\ \tb} \HOB
 & = &
(  \dab  -  \HB  \pa  ) \HZB \, - \,
(\pab - \HZB \pa ) \HB
\nonumber
\end{eqnarray}
and
\begin{eqnarray}
\label{tau}
\tau & = &
( \HO  H_{\tb} ^{\ \tb}  +  H_{\tb} ^{\ \th}  H_{\th} ^{\ \tb}
 )^{-1}  \left[  (  \dab  -  \HB  \pa  )
( \HO  \LA  ) +
(  D  -  \HT  \pa  )
( H_{\tb} ^{\ \th}  \LA  )   \right]
\quad
\\
\bar{\tau} & = &
( \HO  H_{\tb} ^{\ \tb}  +  H_{\tb} ^{\ \th}  H_{\th} ^{\ \tb}
 )^{-1}  \left[  (  D  -  \HT  \pa  )
( H_{\tb} ^{\ \tb}  \LB  ) +
(  \dab  -  \HB  \pa  )
( H_{\th} ^{\ \tb}  \LB  )   \right] .
\nonumber
\end{eqnarray}
The determination of the independent fields in the set of equations
(\ref{26}) is best done by linearizing the variables according
to
$H_{\th} ^{\ \th} = 1+ h_{\th} ^{\ \th},
H_{\tb} ^{\ \tb} = 1+ h_{\tb} ^{\ \tb}$ and
$H_a ^{\ b} = h_a ^{\ b}$ otherwise. The conclusion is the one
summarized above.

Let us complete our discussion
of the $z$-sector. The first
of the structure relations (\ref{strr}) yields, amongst others,
the following differential equation:
\begin{equation}
\label{26a}
0 \, = \, (\, D_a - H_a ^{\ z}  \pa \, ) \, (\LA \LB )
\, - \, (\pa H_a ^{\ z})\,
\LA \LB \, - \, H_a ^{\ \tb} \, \tau \, \LB \, - \, H_a ^{\ \th} \,
\LA \, \bt
\ \ \ \ \ \ {\rm for} \ \ a \, = \, \zb , \th , \tb
.
\end{equation}
We note that this equation also holds for $a=z$ if we
write the generic
elements of the Beltrami matrix $M$ of equation (\ref{19})
as $H_a^{\ b}$ so that
$H_z^{\ z} =1$ and
$H_z^{\ \th} = 0 = H_z^{\ \tb}$. The previous
relation can be decomposed in a symmetric way with respect to
$\LA$ and $\LB$ which leads to the {\em integrating factor
equations} (IFEQ's)
\begin{eqnarray}
0 & = &
(\, D_a  -  H_a ^{\ z} \pa   - \frac{1}{2} \, \pa H_a ^{\ z}
 -  V_a ) \, \LA
\,  - \, H_a ^{\ \tb} \, \tau
\nonumber
\\
0 & =&  (\, D_a  - H_a ^{\ z}
\pa   - \frac{1}{2} \, \pa H_a ^{\ z}
 +  V_a ) \, \LB
\,  - \, H_a ^{\ \th} \, \bt
\ \ .
\label{27}
\end{eqnarray}
The latter decomposition introduces a
vector field $V_a$ (with $V_z =0$)
which is to be interpreted
as a connection for the U(1)-symmetry
due to its transformation law under
U(1)-transformations (see next section).
It should be noted that
$V_a$ is not an independent variable, rather it is determined
in terms of the `$H$' by the structure
equations:
\begin{eqnarray}
V_{\th} & = &
\frac{-1}{H_{\th} ^{\ \th}} \ [  D  -  \HT  \pa
\, + \, \frac{1}{2} \, (\pa \HT )  ]  \, H_{\th} ^{\ \th}
\nonumber
\\
V_{\tb} & = &
\frac{1}{H_{\tb} ^{\ \tb}} \ [  \dab  -  \HB  \pa
\, + \, \frac{1}{2} \, (\pa \HB )  ] \,  H_{\tb} ^{\ \tb}
\label{28}
\\
V_{\zb} & = &
\frac{1}{H_{\th} ^{\ \th}} \, \left\{ [  D  -  \HT  \pa
+  \frac{1}{2} \, (\pa \HT )  + V_{\th} ] \, \HZ  -
[ \pab - \HZB \pa + \frac{1}{2} (\pa \HZB ) ] \, \HO
\right\}
\nonumber
\\
& = &
\frac{-1}{H_{\tb} ^{\ \tb}} \, \left\{ [  \dab  -  \HB  \pa
+  \frac{1}{2} \, (\pa \HB )  - V_{\tb} ] \, \Hzbtb -
[ \pab - \HZB \pa + \frac{1}{2} (\pa \HZB ) ] \, \Htbtb
\right\}
.
\nonumber
\end{eqnarray}
By virtue of the relations between the `$H$', the previous
expressions can be rewritten in various other ways, for
instance
\begin{eqnarray}
\label{rew}
- H_{\tb} ^{\ \th} \, V_{\tb} & = &  [  \dab  -  \HB  \pa
\, + \, \frac{1}{2} \, (\pa \HB )  ] \, H_{\tb} ^{\ \th}
\\
H_{\th} ^{\ \tb} \, V_{\th} & = & [  D  -  \HT  \pa
\, + \, \frac{1}{2} \, (\pa \HT )  ] \, H_{\th} ^{\ \tb}
\ \ .
\nonumber
\end{eqnarray}
This finishes our discussion of the $z$-sector.

In the $\zb$-sector, we have
\begin{equation}
\label{26b}
H_z ^{\ \zb} \, = \, ( \dab \, - \, H_{\tb} ^{\ \zb}  \pab  )
H_{\th} ^{\ \zb}
\, + \, (  D \, - \, H_{\th} ^{\ \zb}  \pab  )  H_{\tb} ^{\ \zb}
\ \ ,
\end{equation}
where $H_{\th}^{\ \zb}$ and
$H_{\tb}^{\ \zb}$ satisfy the covariant chirality conditions
\begin{equation}
\label{26c}
( \, D  -  H_{\th} ^{\ \zb}  \pab \, ) \, H_{\th}^{\ \zb}
\  = \ 0 \ =\
(\, \dab  -  H_{\tb} ^{\ \zb}  \pab \, )\, H_{\tb}^{\ \zb}
\ \ .
\nonumber
\end{equation}
The first condition simply relates the component fields of
$H_{\th}^{\ \zb}$ among themselves and the second those of
$H_{\tb}^{\ \zb}$. Thereby, each of these superfields contains one
independent
bosonic and fermionic space-time component.

The factor $\OM$ satisfies the IFEQ's
\begin{equation}
\label{28a}
0 \ = \ ( \, D_a \, - \, H_{a} ^{\ \zb} \, \pab  \, - \, \pab
H_a ^{\ \zb} \, ) \, \OM
\ \ \ \ \ \ \ \ \ \ {\rm for } \ \ \ a = z   , \th , \tb
\ \ \ ,
\end{equation}
the equation for $z$ being a consequence of the ones for $\th$ and $\tb$.

\section{Symmetry transformations}

To deduce the
transformation laws of the basic fields under
infinitesimal superdiffeomorphisms, we proceed as in the $N=0$ and
$N=1$ theories \cite{dg}. In the
course of this process, the U(1)-transformations manifest themselves
in a natural way.

Thus, we start from the ghost vector field
\[
\Xi \cdot \pa \ \equiv \
\Xi^{z} (z , \zb , \th , \tb )\, \pa \ + \
\Xi^{\zb} (z , \zb , \th , \tb )\, \pab \ + \
\Xi^{\th} (z , \zb , \th , \tb )\, D \ + \
\Xi^{\tb} (z , \zb , \th , \tb )\, \dab
\ \ \ ,
\]
which generates
an infinitesimal change of the coordinates $(z, \zb , \th , \tb )$.
Following C.Becchi \cite{cb, ls},
we consider a reparametrization of
the ghosts,
\begin{equation}
\label{31}
\left( \, C^z \, ,\, C^{\zb} \, ,\, C^{\th} \, ,\, C^{\tb}
\, \right) \ = \
\left( \, \Xi^z \, ,\, \Xi^{\zb} \, ,\, \Xi ^{\th} \, ,\,
\Xi ^{\tb} \, \right) \cdot M
\ \ \ ,
\end{equation}
where $M$ denotes the Beltrami matrix introduced
in equation (\ref{19}). Explicitly,
\begin{eqnarray}
C^z & = & \Xi^z \ + \ \Xi^{\zb} \, H_{\zb} ^{\ z} \ + \
\Xi ^{\th} \, \HT \ + \ \Xi ^{\tb} \, \HB
\nonumber
\\
C^{\zb} & = & \Xi^{\zb} \ + \ \Xi^z \, H_z ^{\ \zb} \ + \
\Xi ^{\th} \, H_{\th} ^{\ \zb} \ + \ \Xi ^{\tb} \, H_{\tb} ^{\ \zb}
\nonumber
\\
C^{\th} & = &
\Xi ^{\th} \, \HO \ + \ \Xi ^{\zb} \, \HZ \ + \
\Xi ^{\tb} \, H_{\tb} ^{\ \th}
\label{32}
\\
C^{\tb} & = &
\Xi ^{\tb} \, H_{\tb} ^{\ \tb} \ +\ \Xi ^{\zb} \, H_{\zb} ^{\ \tb} \ +\
\Xi ^{\th} \, H_{\th} ^{\ \tb}
\ \ \ .
\nonumber
\end{eqnarray}
We note that
the U(1)-transformations of the `$H$', eqs.(\ref{22}), induce those
of the `$C$',
\[
(C^z )^{\prime} \, = \, C^z
\ \ \ , \ \ \
(C^{\zb} )^{\prime} \, = \, C^{\zb}
\ \ \ , \ \ \
(C^{\th} )^{\prime} \, = \, {\rm e}^{-K} \, C^{\th}
\ \ \ , \ \ \
(C^{\tb} )^{\prime} \, = \, {\rm e}^{K} \, C^{\tb}
\ \ ,
\]
but, for
the moment being, we will not consider this symmetry and restrict our
attention to the superdiffeomorphisms.

Contraction of the basis 1-forms (\ref{18})
along the vector field $\Xi \cdot \pa$ gives
\begin{eqnarray}
i_{\Xi \cdot \pa} ( e^Z ) & = &  \left[ \, \Xi^z +
\Xi^{\zb}  {H_{\zb}}^z  +  \Xi^{\th}  \HT  +
\Xi^{\tb}  \HB  \, \right]  \LA_{\th}^{\ \TH}  \LB_{\tb} ^{\ \TB}
\nonumber  \\
& = & C^z  \LA_{\th} ^{\ \TH}  \LB_{\tb} ^{\ \TB}
\label{33}
\\
i_{\Xi \cdot \pa} ( e^{\Theta} ) & = & \left[ \,
\Xi^z  +  \Xi^{\zb}  H_{\zb} ^{\ z}  +
\Xi ^{\th} \HT  +  \Xi ^{\tb} \HB \, \right]  \tau_z ^{\ \TH}
 +  \left[ \, \Xi ^{\th}  \HO  +  \Xi ^{\zb} \HZ  +
\Xi ^{\tb}  H_{\tb} ^{\ \th} \, \right]  \LA_{\th} ^{\ \TH}
\nonumber  \\
& = & C^z  \tau_z ^{\ \TH}  +  C^{\th}  \LA_{\th} ^{\ \TH}
\nonumber
\end{eqnarray}
and similarly
\[
i_{\Xi \cdot \pa} ( e^{\TB} ) \ = \
C^z \, \bt_z ^{\ \TB} \, + \, C^{\tb} \, \LB_{\tb} ^{\ \TB}
\ \ \ \ \ , \ \ \ \ \
i_{\Xi \cdot \pa} ( e^{\ZB} ) \ = \
C^{\zb} \, \OM_{\zb} ^{\ \ZB}
\ \ .
\]
Thereby\footnote{
In superspace, the BRS-operator $s$ is supposed to act as an
antiderivation from the right and the ghost-number is added
to the form degree, the Grassmann parity being $s$-inert \cite{geo}.},
\begin{eqnarray*}
s \TH  & = &
i_{\Xi \cdot \pa} \, d \Theta \, =\, i_{\Xi \cdot \pa}  \, e^{\Theta}
\, = \,
C^z  \tau + C^{\th}  \LA    \\
sZ & = & i_{\Xi \cdot \pa} \,
d Z \, = \, i_{\Xi \cdot \pa}  [\, e^Z  -
\frac{1}{2} \, \TB  e^{\TH}   -
\frac{1}{2} \, \TH  e^{\TB}  \, ]
\, = \, C^z  \LA  \LB
 -  \frac{1}{2} \, \TB  (  s \TH  )
 -  \frac{1}{2} \, \TH  (  s \TB  )
\end{eqnarray*}
and analogously
\[
s\TB \ = \ C^z \, \bt \ + \ C^{\tb} \, \LB
\ \ \ \ \ , \ \ \ \ \
s\ZB \ = \ C^{\zb} \, \OM
 \ \ .
\]
From the nilpotency of the $s$-operation,
$0 = s^2 Z =  s^2 \ZB = s^2 \TH = s^2 \TB$, we now deduce
\begin{eqnarray}
s C^{z} & = & -\, C^z \, (\LA \LB )^{-1} \,
\left[ \, s(\LA \LB ) \, - \, \, C^{\tb} \, \LB \, \tau \, - \,
C^{\th} \, \LA \, \bt
\, \right] \ - \ C^{\th} \, C^{\tb}
\nonumber
\\
s C^{\zb} & = & -\, C^{\zb}  \, \OM ^{-1} \,  \left[ \, s\OM \,
\right]
\nonumber
\\
s C^{\th} & = & - \,  \LA  ^{-1}  \, \left[ \, (sC^z ) \, \tau
\, + \, C^z \, (s\tau ) \,  + \,  C^{\th} \,
(s \LA ) \, \right]
\label{34}
\\
s C^{\tb} & = & - \,  \LB  ^{-1}  \, \left[ \, (sC^z ) \, \bt
\, + \, C^z \, (s\bt ) \, + \, C^{\tb} \,
(s \LB ) \, \right]
\ \ \ .
\nonumber
\end{eqnarray}
The transformation laws of the
integrating factors and Beltrami coefficients follow
by evaluating in two different ways the variations of the
differentials $dZ, d\ZB, d\TH , d\TB$; for instance\footnote{For
the action of the exterior differential
$d$ on ghost fields, see reference \cite{geo}.},
\[
s(d \Theta ) \ = \ -d (s \Theta ) \ = \
+ [ \, e^z \pa \ + \ e^{\zb} \pab \ + \ e^{\th} D
\ + \ e^{\tb} \dab \, ] \, [ \,
C^z \, \tau \, + \,  C^{\th} \, \LA \, ]
\]
and
\begin{eqnarray*}
s(d \Theta ) =  s e^{\Theta} & = &
\left[  e^z  +  e^{\zb} \, \HZB  +  e^{\th} \, \HT  +
e^{\tb} \, \HB  \right]  s\tau  +
\left[  e^{\zb} \, s\HZB  +  e^{\th} \, s\HT  +
e^{\tb} \, s\HB  \right]  \tau
\\
 & & +
\left[ e^{\th} \, \HO  + e^{\zb} \, \HZ  +  e^{\tb} \, \HOB
\right]  s \LA  +
\left[ e^{\th} s\HO   +  e^{\zb} s\HZ
+  e^{\tb}  s\HOB  \right]  \LA
\end{eqnarray*}
lead to the variations of $\tau$ and $\HO , {H_{\zb}}^{\th} ,
{H_{\tb}}^{\th}$.
More explicitly, comparison of the
coefficients of $e^z$ in both expressions for
$s(d\TH)$ yields
\begin{eqnarray}
\label{35a}
s \tau & = &  \pa  \, ( \, C^z  \tau  +   C^{\th}
\LA \, )
\\
s \bt & = &  \pa  \, ( \, C^z  \bt  +   C^{\tb}
\LB \, )
\ \ ,
\nonumber
\end{eqnarray}
where the second equation follows
from $s(d\TB)$
by the same lines of reasoning.
From the coefficients of $e^z$ in $s(dZ)$, one finds
\begin{equation}
s \, (\LA \LB ) \ = \  \pa \, ( C^z  \LA \LB )
\ + \  C^{\tb} \, \LB \, \tau
\ + \  C^{\th} \, \LA \, \bt
\ \ \ .
\end{equation}
In analogy to eqs.(\ref{26a})(\ref{27}), we decompose
this variation in a symmetric way,
\begin{eqnarray}
\label{35}
s \LA & = & C^z \, \pa \LA  \ + \ \frac{1}{2} \  (\pa C^z) \, \LA
\ + \  C^{\tb} \, \tau \ + \ K\, \LA
\\
s \LB & = & C^z \, \pa \LB  \ + \ \frac{1}{2} \  (\pa C^z) \, \LB
\ + \  C^{\th} \, \bt \ - \ K\, \LB
\ \ \ ,
\nonumber
\end{eqnarray}
where $K$ denotes a ghost superfield. The $K$-terms
which naturally appear in this decomposition
represent an infinitesimal version of the
U(1)-symmetry (\ref{22}). The variation of the $K$-parameter
follows from the
requirement that the $s$-operator is nilpotent:
\begin{equation}
s K \, = \, - \left[ \, C^z  \pa  K  -   \frac{1}{2} \,
 C^{\th}  (\pa C^{\tb} ) +  \frac{1}{2} \, C^{\tb}
(\pa C^{\th} ) \, \right]
\ \ .
\end{equation}
By
substituting the expressions (\ref{35a})-(\ref{35}) into eqs.(\ref{34}),
we get
\begin{eqnarray}
s C^z & = & -  \left[ \, C^z  \pa C^z  +  C^{\th}
C^{\tb} \, \right]
\nonumber
\\
s C^{\th} & = & - \left[ \, C^z  \pa  C^{\th}  +  \frac{1}{2}
\, C^{\th}  (\pa C^z )  -  K  C^{\th}  \, \right]
\label{36}
\\
s C^{\tb} & = & - \left[ \, C^z  \pa  C^{\tb}  +  \frac{1}{2}
\, C^{\tb}  (\pa C^z )  +  K  C^{\tb}  \, \right]
\ \ .
\nonumber
\end{eqnarray}

The variations of the Beltrami coefficients
follow by taking into account the previous relations, the
structure equations and eqs.(\ref{27}) where the vector field $V_a$
was introduced.
They take the form
\begin{eqnarray}
\label{37}
sH_a ^{\ z} & = &
(\, D_a   -  H_a ^{\ z} \pa
 +    \pa H_a ^{\ z}  \, )\, C^z  -  H_a ^{\ \th}
C^{\tb}  -  H_a ^{\ \tb}
C^{\th}
\\
s H_a ^{\ \th} & = & ( \, D_a  -  H_a ^{\ z}   \pa
+  \frac{1}{2} \,  \pa H_a ^{\ z}  +  V_a \, ) \, C^{\th}
 +  C^z   \pa H_a ^{\ \th}   -
\frac{1}{2} \,
H_a ^{\ \th}  ( \pa C^z )   -  H_a ^{\ \th}  K
\nonumber
\\
s H_a ^{\ \tb} & = & ( \, D_a  -  H_a ^{\ z}   \pa
+  \frac{1}{2} \,  \pa H_a ^{\ z}  -  V_a \, )  \, C^{\tb}
+  C^z   \pa H_a ^{\ \tb}   -
\frac{1}{2} \,
H_a ^{\ \tb}  ( \pa C^z )  +  H_a ^{\ \tb}  K .
\nonumber
\end{eqnarray}
Finally, the variation of $V_a$ follows by requiring the
nilpotency of the $s$-operations (\ref{37}):
\begin{equation}
\label{38a}
sV_a  =  C^z  \pa V_a  +  \frac{1}{2} \, H_a ^{\ \th}
\pa C^{\tb}
 -  \frac{1}{2} \, (\pa H_a ^{\ \th}) C^{\tb}
 -  \frac{1}{2} \, H_a ^{\ \tb}
\pa C^{\th}
+ \frac{1}{2} \, (\pa H_a ^{\ \tb} ) C^{\th} +
( D_a  -  H_a ^{\ z}  \pa  )  K .
\end{equation}
Equivalently, this transformation law
can be deduced from the variations of the
`$H$' since $V_a$ depends on these variables according to
equations (\ref{28}).
The derivative of $K$ in the
variation (\ref{38a})
confirms the interpretation of $V_a$
as a gauge field for the $U(1)$-symmetry.

In the $\zb$-sector, the same procedure leads to the following
results:
\begin{eqnarray}
s H_a ^{\ \zb} & = & ( \, D_a  -  H_a ^{\ \zb}   \pab
+   \pab H_a ^{\ \zb}  \, )  C^{\zb}
\ \ \ \ \ \ \ \ {\rm for} \ \  a \, =\,  z , \th , \tb
\nonumber
\\
s C^{\zb} & = & -  [ \, C^{\zb}  \pab C^{\zb} \, ]
\label{35b}
\\
s \OM & = & C^{\zb}  \pab \OM  + (\pab C^{\zb})  \OM
\nonumber
\ \ .
\end{eqnarray}
Altogether, the
number of symmetry parameters and independent space-time fields
coincide and the correspondence between them is given by
\begin{equation}
\begin{array}{cccccc}
C^{z} & C^{\th} & C^{\tb} & K &  ; & C^{\zb} \\
H_{\zb}^{\ z} & H_{\tb}^{\ z} & H_{\th}^{\ z} &
H_{\th}^{\ \th} / H_{\tb}^{\ \tb}  & ; &
H_{\th}^{\ \zb} , H_{\tb}^{\ \zb} \ .
\end{array}
\end{equation}
Here, the superfields
$H_{\th}^{\ \zb}$ and $H_{\tb}^{\ \zb}$
are constrained by chirality-type
conditions which reduce the number of their components
by a factor 1/2.

We note that
the {\em holomorphic factorization} is manifestly
realized for the $s$-variations (\ref{35a})-(\ref{35b})
which have explicitly been verified to be nilpotent.
The underlying {\em symmetry group} is the semi-direct product
of superdiffeomorphisms and $U(1)$ transformations:
this fact is best seen by rewriting the infinitesimal
transformations of the ghost fields in terms of the ghost vector
field $\Xi \cdot \pa \,$,
\begin{eqnarray}
s \, ( \Xi \cdot \pa ) & = & - {1 \over 2} \;
[ \, \Xi \cdot \pa \, , \, \Xi \cdot \pa \, ]
\nonumber
\\
s \hat K  & = &
- \, (\Xi \cdot \pa) \, \hat K
 \ \ .
\end{eqnarray}
Here, $[ \ , \ ]$ denotes the graded Lie bracket and
$\hat K = K - i_{\Xi \cdot \pa} V$
is a reparametrization of $K$ involving the
the $U(1)$ gauge field $V= e^a V_a$.
More explicitly, we have
\begin{eqnarray}
s \Xi^z  & = & - \left[
\, (\Xi \cdot \pa ) \, \Xi^z \, - \, \Xi^{\th} \, \Xi^{\tb} \, \right]
\\
s \Xi^a  & = &
- \, (\Xi \cdot \pa) \, \Xi^a \qquad \qquad \qquad
{\rm for} \ \; a= \zb, \th,\tb
\ \ ,
\nonumber
\end{eqnarray}
where the quadratic term
$\Xi^{\th} \Xi^{\tb}$ is due to the fact that the $\Xi^a$ are the
vector components with respect to the canonical tangent space basis
$(D_a)$ rather than the coordinate basis $(\pa_a)$.

Equations (\ref{36})(\ref{35b}) and some of the
variations (\ref{37})-(\ref{38a}) involve only space-time derivatives
and can be projected to component field expressions in a
straightforward way \cite{bbg, dg}. From the definitions
\begin{eqnarray}
\label{39}
{H_{\zb}}^z  \vert & \equiv & {\mu_{\zb}}^z   \ \ \ \ \  ,
\ \ \ \ \ \HZ  \vert \ \equiv \  {\alpha_{\zb}}^{\th}
\\
H_{z} ^{\ \zb} \vert & \equiv & \bar{\mu}_{z} ^{\ \zb}\ \ \ \ \  ,
\ \ \ \ \ H_{\zb} ^{\ \tb}
\vert \ \equiv \  \bar{\alpha}_{\zb} ^{\ \tb}
\ \ \ \ \ , \ \ \ \ \
V_{\zb} \vert \ \equiv \ \bar{v}_{\zb}
\nonumber
\end{eqnarray}
and
\begin{eqnarray}
C^z  \vert & \equiv & c^z \ \equiv \ \xi^z \, + \, \xi^{\zb} \,
{\mu_{\zb}}^z \ \ \ \ , \ \ \ \
C^{\th}  \vert \ \equiv \  \epsilon^{\th}
\ \equiv \ \xi^{\th} \, + \, \xi^{\zb} \, \alpha_{\zb} ^{\ \th}
\nonumber
\\
C^{\zb} \vert & \equiv & \bar{c} ^{\zb}
\ \equiv \ \xi^{\zb} \, + \, \xi^z \,
\bar{\mu}_z ^{\ \zb} \ \ \ \ , \ \ \ \
C^{\tb}  \vert \ \equiv \  \bar{\epsilon} ^{\tb}
\ \equiv \ \xi^{\tb} \, + \, \xi^{\zb} \, \bar{\alpha}_{\zb} ^{\ \tb}
\label{40}
\\
K \vert & \equiv &  k
\; \ \equiv \ \hat k  \; + \, \xi^{\zb} \, \bar{v}_{\zb}
\ \ ,
\nonumber
\end{eqnarray}
we obtain the symmetry algebra of the ordinary Beltrami differentials
($\mu , \bar{\mu}$), of their fermionic partners (the Beltraminos
$\alpha, \bar{\alpha}$) and of the vector $\bar{v}$ :
\begin{eqnarray}
s \mu & = &  ( \, \pab - \mu \, \pa  +
 \pa \mu \, ) \, c -  \bar{\alpha} \, \epsilon -
\alpha \, \bar{\epsilon}
\nonumber
\\
s \alpha & = &
( \, \pab  -   \mu \, \pa    +   \frac{1}{2} \,
\pa \mu   +  \bar v \, ) \, \epsilon
  +  c \,  \pa \alpha    +  \frac{1}{2}\,
\alpha \, \pa c   +   k \, \alpha
\label{41}
\\
s \bar{\alpha} & = &
( \, \pab   -   \mu \, \pa    +   \frac{1}{2}\,
\pa \mu   -  \bar v \, ) \, \bar{\epsilon}   +  c \,  \pa
\bar{\alpha}   +   \frac{1}{2} \, \bar{\alpha} \,
\, \pa c   -   k \, \bar{\alpha}
\nonumber
\\
s \bar{v} & = &
c\, \pa \bar{v}   +   \frac{1}{2} \, \alpha \, \pa \bar{\epsilon}   -
\frac{1}{2} \,
\bar{\epsilon} \, \pa \alpha
  -   \frac{1}{2} \,
\bar{\alpha} \, \pa \epsilon
  +   \frac{1}{2} \,
\epsilon \, \pa \bar{\alpha}
  -    (\, \pab    -    \mu \, \pa \, ) \, k
\nonumber
\\
\nonumber
\\
sc & = & c \,  \pa c    +    \epsilon \, \bar{\epsilon}
\nonumber
\\
s \epsilon & = & c \,  \pa \epsilon
  -   \frac{1}{2} \, \epsilon
\, \pa c     +   k \, \epsilon
\nonumber
\\
s \bar{\epsilon} &=& c \,  \pa \bar{\epsilon}
  -   \frac{1}{2} \, \bar{\epsilon}
\, \pa c     -   k \, \bar{\epsilon}
\nonumber
\\
sk & = & c\, \pa   k
  +   \frac{1}{2} \, \epsilon  \, \pa \bar{\epsilon}
  -   \frac{1}{2} \, \bar{\epsilon}  \, \pa \epsilon
\nonumber
\end{eqnarray}
and, for the $\zb$-sector,
\begin{eqnarray}
s \bar{\mu} & = &  ( \, \pa   -   \bar{\mu} \, \pab    +
 \pab \bar{\mu} \, ) \, \bar{c}
\label{42}
\\
s\bar{c} & = & \bar{c} \,  \pab \bar{c}
\ \ .
\nonumber
\end{eqnarray}
Thus,
the holomorphic factorization remains manifestly realized
at the component field level\footnote{
In equations (\ref{41})(\ref{42}),
$s$ is supposed to act from the
left as usual in component field formalism and the graduation is
given by the sum of the ghost-number and the Grassmann parity;
the signs following from the superspace algebra have been
modified so as to ensure nilpotency of the $s$-operation
with these conventions.}.

\section{Scalar superfields}

In
(2,0) supersymmetry, ordinary scalar fields $X^i (z, \zb)$ generalize to
complex superfields ${\cal X}^i , \, \bar{{\cal X}} ^{\bar{\imath}}
= ({\cal X}^i)^{\ast}$
satisfying the (anti-) chirality conditions
\begin{equation}
\label{50}
D_{\TB} {\cal X}^i \, = \, 0 \, = \,
D_{\TH} \bar{{\cal X}} ^{\bar{\imath}}
\ \ \ .
\end{equation}
The coupling of such fields to
a superconformal class of metrics
on the SRS ${\bf S\Sigma}$
is described by a sigma-model action
\cite{bmg, evo}:
\begin{eqnarray}
S_{inv} [ {\cal X}, \bar{{\cal X}}  ]  & = &  -{i \over 2} \,
\int_{\bf {S\Sigma}}d^4 Z \
[ \, K_j ({\cal X}, \bar{{\cal X}} ) \,
\pa_{\ZB}
{\cal X}^j  \; - \;
\bar K _{\bar \jmath} ({\cal X}, \bar{{\cal X}} ) \,
\pa_{\ZB}
\bar{{\cal X}}^{\bar \jmath} \, ]
\nonumber
\\
& = &
 -{i \over 2} \,
\int_{\bf {S\Sigma}}d^4 Z \
K_j ({\cal X}, \bar{{\cal X}} ) \,
\pa_{\ZB}
{\cal X}^j  \; + \; {\rm h.c.}
\ \ .
\label{51}
\end{eqnarray}
Here,
$d^4 Z = dZ \, d\bar{Z} \, d\Theta \, d\bar{\Theta}$ and
$K_j$ denotes an arbitrary complex function (and
$\bar K _{\bar \jmath} = (K_j)^{\ast}$ in the Minkowskian setting).
The functional (\ref{51})
is invariant under superconformal changes of coordinates
for which the measure $d^4Z$ transforms with
$(D_{\TH} \TH ^{\prime} )^{-1} \,
(D_{\TB} \TB ^{\prime} )^{-1}$, i.e. the Berezinian associated to the
superconformal transformation (\ref{4})-(\ref{3c}).

We now rewrite the expression (\ref{51})
in terms of the reference coordinates $(z,\zb,\th,\tb)$ by means of
Beltrami superfields. The passage from the small to the capital
coordinates reads
\begin{equation}
\label{52}
\left( \begin{array}{c}
\pa_Z \\ \pa_{\ZB} \\ D_{\TH} \\ D_{\TB}
\end{array} \right)
\ = \ Q^{-1} \ M^{-1} \
\left( \begin{array}{c}
\pa \\ \pab \\ D \\ \bar{D}
\end{array} \right)
\end{equation}
and the Berezinian of this change of variables is
\begin{equation}
\label{53}
\left| \frac {\pa (Z,\ZB,\TH,\TB)}{\pa (z,\zb,\th,\tb)} \right|
\ =\ {\rm sdet}\, (M\,Q) \ = \ \OM \, {\rm sdet}\, M \ \ .
\end{equation}
The inverse of $Q$ is easily determined:
\begin{equation}
\label{54}
Q^{-1} \ = \ \left( \begin{array}{cccc}
\LA^{-1}\LB^{-1} & 0 & -\LA^{-2}\LB^{-1} \tau & -\LA^{-1}\LB^{-2}
\bar{\tau} \\
0 & \OM^{-1} & 0 & 0 \\
0 & 0 & \LA^{-1} & 0 \\
0 & 0 & 0 & \LB^{-1} \\
\end{array} \right)
\ \ .
\end{equation}
In order to calculate sdet\,$M$ and $M^{-1}$, we
decompose $M$ according to
\begin{equation}
\label{55}
M =
\left( \begin{array}{cccc}
1 & 0 & 0 & 0 \\
0 & 1 & 0 & 0 \\
\htz & \htzb & 1 & 0 \\
\htbz & \htbzb & 0 & 1 \\
\end{array} \right)
\left( \begin{array}{cccc}
1 & H_{z}^{\ \zb} & 0 & 0 \\
\HZB & 1 & 0 & 0 \\
0 & 0 & \htt & \httb \\
0 & 0 & \htbt & \htbtb \\
\end{array} \right)
\left( \begin{array}{cccc}
1 & 0 & \hzt & \hztb \\
0 & 1 & \hzbt & \hzbtb \\
0 & 0 & 1 & 0 \\
0 & 0 & 0 & 1 \\
\end{array} \right)
\ .
\end{equation}
The explicit expressions for the `$h$' are
\begin{equation}
\label{delt}
\begin{array}{lcl}
\htz \ = \  \DE^{-1}(\HT-  H_{\th}^{\ \zb} \HZB ) & ,
&\htzb \ = \ \DE^{-1}(H_{\th}^{\ \zb}- \HT H_{z}^{\ \zb} ) \\
\htbz \ = \ \DE^{-1}(\HB- H_{\tb}^{\ \zb} \HZB ) & ,
&\htbzb \ = \ \DE^{-1}(H_{\tb}^{\ \zb}- \HB H_{z}^{\ \zb} )\\
\htt \ = \ \HO-\htzb\HZ & ,
&\httb \ = \ H_{\th}^{\ \tb}-\htzb H_{\zb}^{\ \tb}\\
\htbt \ = \ \HOB-\htbzb\HZ & ,
&\htbtb \ = \ H_{\tb}^{\ \tb}-\htbzb H_{\zb}^{\ \tb}\\
\hzt \ = \ -\DE^{-1}H_{z}^{\ \zb}\HZ & ,
&\hztb \ = \ -\DE^{-1}H_{z}^{\ \zb}H_{\zb}^{\ \tb}\\
\hzbt \ = \ \DE^{-1}\HZ & ,&\hzbtb \ = \ \DE^{-1}H_{\zb}^{\ \tb}
\ \ ,
\end{array}
\end{equation}
where $\Delta = 1 - H_z^{\ \zb}
H_{\zb}^{\ z}$.
It follows that sdet$\, M =\Delta/h$ with $h=\htt\htbtb-\htbt\httb$
and that
\[
M^{-1} =
\left( \begin{array}{cccc}
1 & 0 & -\hzt & -\hztb \\
0 & 1 & -\hzbt & -\hzbtb \\
0 & 0 & 1 & 0 \\
0 & 0 & 0 & 1
\end{array} \right)
\qquad \qquad \qquad \qquad \qquad
\qquad \qquad \qquad\qquad \qquad
\]
\[
\qquad \qquad
\times
\left( \begin{array}{cccc}
1/\DE & -H_{z}^{\ \zb}/\DE & 0 & 0 \\
-\HZB/\DE & 1/\DE & 0 & 0 \\
0 & 0 & \htbtb/h & -\httb/h \\
0 & 0 & -\htbt/h & \htt/h
\end{array} \right)
\left( \begin{array}{cccc}
1 & 0 & 0 & 0 \\
0 & 1 & 0 & 0 \\
-\htz & -\htzb & 1 & 0 \\
-\htbz & -\htbzb & 0 & 1
\end{array} \right)
.
\]
From these results and equation (\ref{52}),
we can derive explicit expressions for
$\pa_Z, \pa_{\ZB}, D_{\TH}, D_{\TB}$ which imply
\begin{eqnarray}
D_{\TB} {\cal X}^i  =  0 &  & \Leftrightarrow \ \
h_{\th} ^{\ \th} ( \dab - h_{\tb}^{\ z} \pa - h_{\tb}^{\ \zb} \pab)
{\cal X}^i \, = \,
h_{\tb} ^{\ \th} ( D - h_{\th}^{\ z} \pa - h_{\th}^{\ \zb} \pab)
{\cal X}^i  \quad
\nonumber
\\
D_{\TH} \bar{{\cal X}} ^{\bar \imath} = 0 &  & \Leftrightarrow \ \
h_{\tb} ^{\ \tb} ( D - h_{\th}^{\ z} \pa - h_{\th}^{\ \zb} \pab)
\bar{{\cal X}} ^{\bar \imath} \, = \,
h_{\th} ^{\ \tb} ( \dab - h_{\tb}^{\ z} \pa - h_{\tb}^{\ \zb} \pab)
\bar{{\cal X}} ^{\bar \imath}
. \quad
\end{eqnarray}
Furthermore,
by substituting $\pa_{\ZB}$ into the action (\ref{51})
and taking into account the last relation for ${\cal X}^i$,
one obtains the {\em final result}
\begin{equation}
\label{57}
S_{inv} [ {\cal X}, \bar{{\cal X}} ]  =  -{i \over 2} \,
\int_{\bf {S\Sigma}} d^{4}z \,
K_j ({\cal X} ,  \bar{{\cal X}} )
\, \bar{\nabla} {\cal X}^j  \, + \, {\rm h.c.}
\ \ ,
\end{equation}
where
$d^{4}z \, = \, dz \, d\bar{z} \, d\theta \, d\bar{\theta}$ and
\begin{equation}
\label{57a}
\bar{\nabla} =
\ds{1 \over h}  ( \pab-\HZB\pa)
+  \ds{1 \over h^2}
H_{\zb}^{\ \th} \left[ \httb
(\bar{D}-\htbz\pa-\htbzb\pab)  -  h_{\tb}^{\ \tb}
(D-\htz\pa-\htzb\pab)
\right]
\ .
\end{equation}

\section{Intermediate coordinates}

If we disregard the complex conjugation relating $z$ and $\zb$,
we can introduce the so-called
intermediate or `tilde' coordinates \cite{dg} by
\[
(z, \zb , \th, \tb ) \ \stackrel{M_1 Q_1}{\longrightarrow} \
(\zt, \tilde{\zb} , \tilde{\th} , \tilde{\tb} ) = (Z, \zb , \TH , \TB )
\ \stackrel{M_2 Q_2}{\longrightarrow}
\ (Z, \ZB , \TH , \TB )
\ \ .
\]
The matrix $M_1 Q_1$ describing the passage from
$(z, \zb , \th, \tb )$ to
$(\zt , \tilde{\zb} ,
\tilde{\th} , \tilde{\tb} )$ is easy to invert: in analogy
to eq.(\ref{52}), we thus obtain the tilde derivatives
\begin{eqnarray}
\dt & = & \frac{1}{\LA H} \ \left[ H_{\tb}^{\ \tb}
(D - \HT \pa ) - H_{\th}^{\ \tb} (\dab - \HB \pa ) \right]
\nonumber
\\
\tilde{\dab}  & = &
\frac{1}{\LB H} \ \left[ H_{\th}^{\ \th}
(\dab  - \HB \pa ) - H_{\tb}^{\ \th} ( D - \HT \pa ) \right]
\\
\tilde{\pa} & = &
\frac{1}{\LA \LB} \ \left[ \pa - \tau \dt - \bar{\tau} \tilde{\dab}
\right]
\nonumber
\\
\pabt & = & (\pab - \HZB \pa) - \LA H_{\zb} ^{\ \th} \dt
- \LB H_{\zb} ^{\ \tb} \tilde{\dab}
\nonumber
\ \ ,
\end{eqnarray}
where $H =H_{\th} ^{\ \th} H_{\tb} ^{\ \tb} -
H_{\tb} ^{\ \th} H_{\th} ^{\ \tb}$.
For later reference, we note that
${\rm sdet} \, (M_1 Q_1) = H^{-1}$.

For the passage from the tilde to the capital coordinates,
we have
\begin{eqnarray}
D_{\TH} & = & \tilde{D} - k_{\th} ^{\ \zb} \tilde{\pab}
\qquad , \qquad
\pa_Z \ = \ \tilde{\pa}  - k_z ^{\ \zb} \tilde{\pab}
\nonumber
\\
D_{\TB} & = & \tilde{\dab} - k_{\tb} ^{\ \zb} \tilde{\pab}
\qquad , \qquad
\pa_{\ZB} \ = \ \Omega^{-1}  \tilde{\pab}
\ \ ,
\nonumber
\end{eqnarray}
where the explicit form of the `$k$' in terms of the `$H$'
and $\LA, \LB$ follows from the condition
$MQ= (M_1 Q_1)(M_2 Q_2)$.

As a first application of the tilde coordinates, we prove
that the solutions of the IFEQ's (\ref{27}) for $\LA$ and $\LB$
are determined up to superconformal transformations of the
capital coordinates, i.e. up to the rescalings (\ref{17}).
In fact, substitution of the expressions (\ref{tau}) for $\tau$ and
$\bar{\tau}$ into the IFEQ's (\ref{27})
shows that the homogenous equations associated to the IFEQ's
can be rewritten as
\begin{eqnarray}
0 & = & \dt \, {\rm ln} \, \LA =
\tilde{\pab}
\, {\rm ln} \, \LA
\qquad \Longrightarrow \qquad
0  =  D_{\TH} \, {\rm ln} \, \LA =
\pa_{\ZB}
\, {\rm ln} \, \LA
\\
\nonumber
0 & = & \dabt \, {\rm ln} \, \LB =
\tilde{\pab}
\, {\rm ln} \, \LB
\qquad \Longrightarrow \qquad
0  =  D_{\TB} \, {\rm ln} \, \LB =
\pa_{\ZB}
\, {\rm ln} \, \LB
\ \ .
\end{eqnarray}
Henceforth, the solutions $\LA , \LB$
of the IFEQ's are determined up to
the rescalings
\begin{eqnarray*}
\LA^{\prime} & = & {\rm e}^{\, f(Z, \TH , \TB )} \LA
\qquad  {\rm with} \quad D_{\TH} f = 0
\\
\LB^{\prime} & = & {\rm e}^{\, g(Z, \TH ,\TB )} \LB
\qquad  {\rm with} \quad D_{\TB} g = 0
\ \ ,
\end{eqnarray*}
which correspond precisely to the superconformal transformations
(\ref{17}).

Another application of the tilde coordinates consists
of the determination of anomalies and effective actions and will be
presented in section 3.8.

Since the $z$- and $\zb$-sectors do not play a symmetric r\^ole
in the (2,0)-theory, we can introduce a second set of
intermediate coordinates which will be referred to as `hat' coordinates:
\[
(z, \zb , \th, \tb ) \ \stackrel{\hat M_1 \hat Q_1}{\longrightarrow} \
(\hat{z} , \hat{\zb} , \hat{\th} , \hat{\tb} ) = (z, \ZB , \th , \tb )
\ \stackrel{\hat M_2 \hat Q_2}{\longrightarrow}
\ (Z, \ZB , \TH , \TB )
\ \ .
\]
Using the hat derivatives
\begin{eqnarray}
\label{coz}
\hat D  & = & D - H_{\th} ^{\ \zb} \pab
\qquad , \qquad
\hat{\pa} \ = \ \pa - H_z^{\ \zb} \pab
\\
\hat{\dab}  & = &\dab - H_{\tb}^{\ \zb} \pab
\qquad , \qquad
\hat{\pab} \ = \ \Omega ^{-1} \pab
\ \ ,
\nonumber
\end{eqnarray}
one proves that the ambiguity of the
solutions of the IFEQ's for $\Omega$ coincides
with superconformal rescalings.

By construction, the derivatives (\ref{coz}) satisfy the same algebra
as the basic differential operators $(\pa, \pab, D , \dab)$,
in particular,
\begin{equation}
\label{coa}
\{ \hat D , \hat{\dab} \} = \hat{\pa}
\qquad , \qquad
\hat D ^2 = 0 = \hat{\dab} ^2
\qquad , \qquad
[ \hat D , \hat{\pa} ] = 0 = [ \hat{\dab} ,  \hat{\pa}]
\ \ .
\end{equation}
By virtue of these derivatives, the solution (\ref{26b})(\ref{26c})
of the structure relations in the $\zb$-sector can be rewritten
in the compact form
\begin{equation}
\label{soa}
H_z^{\ \zb} =
\hat{\dab} H_{\th} ^{\ \zb} +
\hat D  H_{\tb} ^{\ \zb}
\qquad  ,  \qquad
\hat D  H_{\th} ^{\ \zb}= 0 =
\hat{\dab} H_{\tb} ^{\ \zb}
\ \ ,
\end{equation}
which equations will be further exploited in section 3.9.

\section{Restriction of the geometry}

In the study of the $N=1$ theory, it was noted that the choice
$\HT =0$ is invariant under superconformal
transformations so that are no global obstructions for
restricting the geometry by this condition. In fact, this
choice greatly simplifies expressions involving Beltrami superfields
and it might even be
compulsory for the study of specific problems \cite{dg1, cco}.
As for the physical interpretation,
the elimination
of $\HT$ simply amounts to disregarding some
pure gauge fields.

In the following, we introduce the $(2,0)$-analogon of the
$N=1$ condition
$\HT =0$.
In the present case, we have a greater freedom to impose
conditions: this can be illustrated by the fact that a restriction
of the form
$DC^z =0$ on the superdiffeomorphism parameter $C^z$
does not imply $\pa C^z =0$ (i.e. a restricted space-time
dependence of $C^z$) as it does in the $N=1$ theory.
The analogon of the $N=1$ restriction of the geometry is defined by
the relations
\begin{equation}
\label{h1}
\HT =0 = \HB
\qquad {\rm and} \qquad
H_{\th}^{\ \th} \, / \,
H_{\tb}^{\ \tb}  = 1
\end{equation}
in the $z$-sector and
\begin{equation}
\label{h2}
H_{\tb}^{\ \zb} = 0
\end{equation}
in the $\zb$-sector. (The latter condition could also be replaced by
$H_{\th}^{\ \zb} =0$ since equations (\ref{app}) following
from the structure relations in the $\zb$-sector are
symmetric with respect to $\th$ and $\tb$.)
Conditions (\ref{h1}) and (\ref{h2}) are compatible
with the superconformal transformation laws (\ref{24}).

In the remainder of the text, we will consider the geometry constrained
by equations (\ref{h1}) and (\ref{h2}) which will be
referred to as the
{\em restricted geometry}. In this case, there is one
unconstrained Beltrami superfield in the $z$-sector, namely
$\HZB$, and one superfield in the $\zb$-sector, namely
$H_{\th}^{\ \zb}$, subject to the condition
$(D-
H_{\th}^{\ \zb} \pab )
H_{\th}^{\ \zb} =0$.
The relations which hold for the other variables become
\begin{eqnarray}
\nonumber
D\LA \! & = & \! 0 \quad , \quad \tau  =  \dab \LA \quad , \quad
H_{\th}^{\ \th} = 1  \quad , \quad H_{\tb} ^{\ \th} =0 \quad , \quad
\HZ  =  \dab \HZB \\
\nonumber
\dab\LB \! & = & \! 0 \quad , \quad \bar{\tau} = D \LB \quad , \quad
H_{\tb}^{\ \tb} = 1  \quad , \quad H_{\th} ^{\ \tb} =0 \quad , \quad
H_{\zb}^{\ \tb}  =  D \HZB
\\
V_{\th} \! & = & \! 0
\quad , \quad
V_{\tb} =  0
\quad \ \ , \quad \ \,
V_{\zb}  =  {1 \over 2} \, [ D,\dab ] \HZB
\\
&&
\nonumber
\\
\dab \Omega & = & 0 \quad , \quad
H_z^{\ \zb} \, = \, \dab H_{\th} ^{\ \zb}  \quad , \quad
(D -
H_{\th} ^{\ \zb} \pab )
H_{\th} ^{\ \zb} \, = \, 0
\ \ ,
\nonumber
\end{eqnarray}
while
the superconformal transformation laws now read
\begin{eqnarray}
\LA ^{\pr} & = & {\rm e}^w \, \LA
\quad , \quad
\LB ^{\pr} \, = \, {\rm e}^{\bar w} \, \LB
\quad , \quad
H_{\zb^{\prime}} ^{\ \, z^{\prime}} \, = \, {\rm e}^{-w-\bar{w}}
\, (\pab \zb ^{\prime} )^{-1}  \, \HZB
\nonumber
\\
&&
\nonumber
\\
\Omega^{\prime} & = &
\, (\pab \zb ^{\prime} )^{-1} \, \Omega
\quad , \quad
H_{\th^{\prime}} ^{\ \, \zb^{\prime}} \, = \, {\rm e}^{w}
\, (\pab \zb ^{\prime} ) \, H_{\th}^{\ \zb}
\ \ .
\nonumber
\end{eqnarray}
Furthermore,
from (\ref{ana}) and (\ref{13a}), we get the local expressions
\begin{eqnarray*}
\LA & = & D \TH \qquad , \qquad \LB = \dab \TB
\\
& &
\\
\Omega & = & \pab \ZB
\qquad ({\rm as} \ {\rm before})
\ \ .
\end{eqnarray*}
In order to be consistent, we have to require that the
conditions (\ref{h1}) and (\ref{h2}) are invariant under the
BRS transformations. This determines the symmetry parameters $C^{\th},
\, C^{\tb}, \, K$ in terms of $C^z$ and eliminates some
components of $C^{\zb}$:
\begin{eqnarray}
C^{\th} & = & \dab C^z
\quad , \quad
C^{\tb} \, = \, D C^z
\quad , \quad
K \, = \, {1 \over 2} \, [ D, \dab ] C^z
\nonumber
\\
&&
\nonumber
\\
\dab C^{\zb} & = & 0
\ \ .
\end{eqnarray}
The $s$-variations of the basic variables in the $z$-sector then take
the form
\begin{eqnarray}
s \HZB  &=&  [ \, \pab - \HZB \pa - (\dab \HZB ) D - (D \HZB) \dab
+( \pa \HZB) \, ] \, C^z
\nonumber
\\
s \LA  &=&  [ \,  C^z \pa + (DC^z ) \dab \, ] \, \LA \, + \,
(D \dab C^z ) \, \LA
\nonumber
\\
s \LB  &=&  [ \,  C^z \pa + (\dab C^z ) D \, ]  \, \LB \, + \,
(\dab D C^z ) \, \LB
\\
s C^z & = &  - \, [ \,  C^z \pa C^z + (\dab C^z )( DC^z ) \, ]
\ \ ,
\nonumber
\end{eqnarray}
while those in the $\zb$-sector are still given by equations
(\ref{35b}).

Finite superdiffeomorphisms can be discussed
along the lines of the $N=1$ theory \cite{dg}. Here, we only note
that the restriction (\ref{h1})(\ref{h2})
on the geometry reduces the symmetry
group ${\rm sdiff} \, {\bf S \Sigma} \otimes U(1)$
to a subgroup thereof.

\section{Component field expressions}

In the restricted geometry (defined in the previous section),
the basic variables of the $z$-sector
are the superfields $\HZB$ and $C^z$ which have the following
$\th$-expansions:
\begin{eqnarray}
\HZB & = & \mu_{\zb}^{\ z} + \th \, \bar{\alpha}_{\zb}^{\ \tb}
+ \tb \, \alpha_{\zb} ^{\ \th} + \tb \th \, \bar v _{\zb}
\nonumber  \\
\label{cf}
C^z & = & c^z + \th \,  \bar{\epsilon}^{\tb}
+ \tb \,  \epsilon ^{\th}
+ \tb \th \,  k
\ \ .
\end{eqnarray}
Here, the bosonic fields $\mu$ and $\bar v$ are the ordinary
Beltrami coefficient and the $U(1)$ vector while $\alpha$
and $\bar{\alpha}$ represent their fermionic partners,
the Beltraminos. These variables transform under general
coordinate, local supersymmetry and local $U(1)$-transformations
parametrized, respectively, by $c, \epsilon,
\bar{\epsilon}$ and $k$.

The basic variables of the $\zb$-sector are $H_{\th}^{\ \zb}$
and $C^{\zb}$. To discuss their field content, we choose
the {\em WZ-supergauge} in which the only non-vanishing component
fields are
\begin{equation}
\dab H_{\th}^{\ \zb} \! \mid \; = \, \bar{\mu} _z ^{\ \zb}
\qquad {\rm and} \qquad
C^{\zb} \! \mid \; = \, \bar{c} ^{\zb} \quad , \quad
\dab D C^{\zb} \! \mid \; = \, \pa \bar{c} ^{\zb}
\ \ .
\end{equation}
As expected for the (2,0)-supersymmetric theory, the $\zb$-sector
only involves the complex conjugate of $\mu$ and $c$.

In the remainder of this section, we present the component
field results in the WZ-gauge.
For the matter sector, we consider a single superfield
${\cal X}$ (and its complex conjugate $\bar{\cal X}$) and a flat
target space metric ($K_j = \delta_{j \bar{\imath} } \, \bar{{\cal X}}
^{\bar \imath}$).
Henceforth, we only have one complex scalar and two spinor fields as
component fields:
\begin{eqnarray}
{\cal X} \! \mid & \equiv & X \qquad , \qquad
D{\cal X} \! \mid \ \equiv \ \lambda_{\th}
\nonumber  \\
\bar{{\cal X}} \! \mid & \equiv & \bar X \qquad , \qquad
\dab \bar{{\cal X}} \! \mid \ \equiv \ \bar{\lambda} _{\tb}
\label{xcomp}
\ \ .
\end{eqnarray}
For these fields,
the invariant action (\ref{57}) reduces to the following
functional on the Riemann surface $\bf{{\Sigma}}$:
\begin{eqnarray}
i \, S_{inv} &= &
\int_{\bf{\Sigma}} d^2z \ \left\{
\frac{1}{1 - \mu \bar{\mu} } \ \left[
(\pab - \mu \pa ) X \,
(\pa - \bar{\mu} \pab ) \bar X
\right. \right.
\\
& & \qquad \qquad \qquad \left.
- \alpha \lambda
(\pa - \bar{\mu} \pab ) \bar X -
\bar{\alpha} \bar{\lambda}
(\pa - \bar{\mu} \pab )  X -  \bar{\mu}
(\alpha \lambda)
(\bar{\alpha} \bar{\lambda}) \right]
\nonumber
\\
& & \qquad \qquad \qquad \qquad \qquad \qquad \qquad \qquad \left.
- \bar{\lambda}
(\pab - \mu \pa - {1 \over 2} \pa \mu - \bar v ) \lambda \right\}
\ \ .
\nonumber
\end{eqnarray}
The $s$-variations of the matter superfields,
$s{\cal X} = (\Xi \cdot \pa ) {\cal X} , \,
s\bar{{\cal X}} = (\Xi \cdot \pa ) \bar{{\cal X}}$ can be projected
to space-time in a straightforward manner: from the definitions
$\Xi^z \! \mid  \, \equiv \, \xi , \,
\Xi^{\zb} \! \mid  \, \equiv \, \bar{\xi} , \,
\Xi^{\th} \! \mid  \, \equiv \, \xi^{\th} , \,
\Xi^{\tb} \! \mid  \, \equiv \, \xi^{\tb}$ and
(\ref{cf})-(\ref{xcomp}),
it follows that
\begin{eqnarray}
sX \! & \! = \! & \! (\xi \cdot \pa ) X + \xi^{\th} \lambda
\quad , \quad
s\lambda \, = \, (\xi \cdot \pa ) \lambda
+{1 \over 2}  (\pa \xi + \mu \pa \bar{\xi} ) \lambda
+ \hat k \lambda + \xi^{\tb} {\cal D} X \quad \ \
\\
s\bar X \! &\! = \! &\! (\xi \cdot \pa ) \bar X + \xi^{\tb} \bar{\lambda}
\quad , \quad
s\bar{\lambda} \; = \; (\xi \cdot \pa ) \bar{\lambda}
+{1 \over 2}  (\pa \xi + \mu \pa \bar{\xi} ) \bar{\lambda}
- \hat k \bar{\lambda} + \xi^{\th} {\cal D} \bar X
\ , \quad \ \
\nonumber
\end{eqnarray}
where we introduced the notation
$\xi \cdot \pa \equiv \xi \pa + \bar{\xi} \pab, \,
\hat k \equiv k - \bar{\xi} \bar v$ and the supercovariant
derivatives
\begin{equation}
{\cal D} X  =
\frac{1}{1 - \mu \bar{\mu} } \ \left[
(\pa - \bar{\mu} \pab ) X + \bar{\mu} \alpha \lambda \right]
\quad , \quad
{\cal D} \bar X  =
\frac{1}{1 - \mu \bar{\mu} } \ \left[
(\pa - \bar{\mu} \pab ) \bar X + \bar{\mu} \bar{\alpha} \bar{\lambda}
\right]
\ .
\end{equation}

\section{Anomalies and effective actions}

For the discussion of the
chirally split form of the superdiffeomorphism anomaly
and of its compensating action, we again
consider the restricted geometry defined in section 3.6.
We follow the procedure developed in reference \cite{dg1}
for the bosonic and $N=1$ supersymmetric cases and we expect that
the results can be extended to the {\em un}restricted geometry at the
expense of technical complications
as in the $N=1$ case.
We will mainly
work on the superplane
${\bf SC}$, but we will also comment on the
generalization to generic compact SRS's.
The results for the $\zb$-sector
are to be discussed in the next section.

The
{\em holomorphically split form of the superdiffeomorphism anomaly} on
the superplane is given in the $z$-sector by
\begin{eqnarray}
\label{an}
{\cal A}^{(z)} [C^z ; \HZB ] & = &
\int_{\bf SC} d^4z \ C^z \, \pa [ D, \dab ] \, \HZB
\\
& = &
{1 \over 2 } \
\int_{\bf C} d^2z \ \left\{ c \pa^3 \mu
+ 2 \epsilon \pa^2 \bar{\alpha}
+ 2 \bar{\epsilon} \pa^2 \alpha
+ 4 k \pa \bar v \right\}
\ \ .
\nonumber
\end{eqnarray}
It satisfies the Wess-Zumino (WZ) consistency condition
$s{\cal A} = 0$.
An expression which is well defined on a generic compact SRS
is obtained by replacing the operator
$\pa [D, \dab ]$ by the superconformally covariant operator
\begin{equation}
{\cal L}_2= \pa [ D, \dab ] + {\cal R} \pa
- (D{\cal R} ) \dab
- (\dab {\cal R} ) D
+ (\pa  {\cal R} )
\label{bol}
\end{equation}
depending on a superprojective connection ${\cal R}$
\cite{ip}; from
$s{\cal R} =0$, it follows that the so-obtained functional
still satisfies the WZ consistency condition.

We note that
our superspace expression for ${\cal A}$ was previously found in
Polyakov's light-cone gauge \cite{xu} and that the corresponding
component field expression coincides with the
result found in
reference \cite{ot}
by differential geometric methods.

If written in terms
of the tilde coordinates,
the {\em Wess-Zumino-Polyakov (WZP) action}
associated to the chirally split superdiffeomorphism
anomaly on ${\bf SC}$
has the form of a free
scalar field action for the integrating factor
\cite{dg1}. Thus, in the present case, it reads
\begin{equation}
\label{wzp}
S^{(z)} _{WZP} [ \HZB ] =
\int_{{\bf SC}} d^4 \zt \ {\rm ln} \, \LB \, (\pabt \, {\rm ln} \, \LA )
\ \ ,
\end{equation}
where the variables
${\rm ln} \, \LA$ and
${\rm ln} \, \LB$ represent (anti-) chiral superfields with respect to
the tilde coordinates: $\dt \, {\rm ln} \, \LA =0 =
\tilde{\dab} \, {\rm ln} \, \LB$. By
rewriting the action in terms of the coordinates $(z, \zb, \th, \tb )$
and applying the $s$-operation, one reproduces the anomaly (\ref{an}):
\begin{eqnarray}
\label{wzpa}
S^{(z)} _{WZP} [ \HZB ] & = &
- \int_{{\bf SC}} d^4z \ \HZB (\pa \, {\rm ln} \, \LB )
\\
s S^{(z)} _{WZP} [ \HZB ] & = &
- {\cal A}^{(z)} [C^z ; \HZB  ]
\ \ .
\nonumber
\end{eqnarray}

The response of the WZP-functional to an infinitesimal variation
of the complex structure $(H_{\zb}^{\ z} \to
H_{\zb}^{\ z} + \delta H_{\zb}^{\ z}$) is given by the super
Schwarzian derivative,
\begin{equation}
\frac{\delta S_{WZP}^{(z)} }{\delta H_{\zb} ^{\ z} }
= {\cal S} (Z , \TH ;  z , \th )
\ \ ,
\end{equation}
the latter being defined by \cite{jc, ar, ip}
\begin{equation}
{\cal S} (Z , \TH ;  z , \th )
= [ D , \dab ] Q - (DQ )(\dab Q) \qquad {\rm with} \quad
Q = {\rm ln} \, D\TH  \, + \, {\rm ln} \, \dab \TB
\ \ .
\end{equation}
The proof of this result proceeds along the lines of reference
\cite{dg1}: it makes use of the IFEQ's for $\LA = D\TH, \,
\LB = \dab \TB$ and of the fact that the functional
(\ref{wzp}) can be rewritten as
\begin{eqnarray}
S^{(z)} _{WZP} [ \HZB ] & = & {1 \over 2}
\int_{{\bf SC}} d^4 \zt \ \left[
\, {\rm ln} \, \LB \; \pabt \, {\rm ln} \, \LA
- {\rm ln} \, \LA \; \pabt \, {\rm ln} \, \LB  \, \right]
\nonumber \\
& = & {1 \over 2}
\int_{{\bf SC}} d^4z \ \left[ \, {\rm ln} \, \LB \ D\dab \HZB -
{\rm ln} \, \LA \ \dab D \HZB \, \right]
\ \ .
\end{eqnarray}

Within the framework of (2,0) supergravity (i.e. the metric approach),
the effective action $S_{WZP}^{(z)}$
represents a chiral gauge expression (see \cite{dg1} and
references therein): in this approach, it rather takes the form
\begin{equation}
S^{(z)} _{WZP} =
- \int_{{\bf SC}} d^4z  \ {\pa \TB \over \dab \TB} \; \dab \HZB
\ \ ,
\end{equation}
which follows from (\ref{wzpa}) by substitution of
$\LB = \dab \TB$.

We note that the
extension of the WZP-action from ${\bf SC}$
to generic super Riemann surfaces has been
discussed for the $N=0$ and $N=1$ cases in references \cite{ls, rz}
and \cite{ak}, respectively.

The {\em anomalous Ward identity} on the superplane reads
\begin{equation}
- \int_{{\bf SC}} d^4z \ (s\HZB)
\frac{\delta Z_c}{\delta \HZB} \, = \, k \,
{\cal A}^{(z)} [C^z ; \HZB  ]
\ \ ,
\end{equation}
where $Z_c$ denotes the vertex functional and $k$ a constant.
By substituting the explicit expression for $s\HZB$ and introducing
the super stress tensor ${\cal T}_{\th \tb} =
\delta Z_c \, / \, \delta \HZB$, the last equation takes
the local form
\begin{equation}
\left[ \pab - \HZB \pa - (\dab \HZB ) D - (D\HZB) \dab - (\pa \HZB)
\right]
{\cal T}_{\th \tb} \, = \, - k \, \pa [D, \dab ] \HZB
\ \ .
\end{equation}
This relation
has previously been derived and discussed in the light-cone gauge
\cite{xu}.
For $k\neq 0$, the redefinition ${\cal T} \to -k {\cal T}$
yields
\[
{\cal L}_2 \HZB = \pab {\cal T}_{\th \tb}
\ \ ,
\]
where ${\cal L}_2$ represents the covariant operator ({\ref{bol})
with ${\cal R} = {\cal T}$.

\section{The $\zb$-sector revisited}

Since the hat derivatives $\hat D$ and $\hat{\dab}$ are nilpotent,
the constraint equations (\ref{soa}), i.e.
$\hat D  H_{\th} ^{\ \zb} = 0 =
\hat{\dab} H_{\tb} ^{\ \zb}$,
can be solved in terms of superfields
$H^{\zb}$ and $\check H ^{\zb}$:
\begin{eqnarray}
H_{\th} ^{\ \zb} & = &
\hat D H^{\zb} \, = \, (D -
H_{\th} ^{\ \zb} \pab ) H^{\zb} \, = \,
\sum_{n=0}^{\infty} ( - \pab H^{\zb} )^n
\ D H^{\zb}
\\
\nonumber
H_{\tb} ^{\ \zb} & = &
\hat{\dab} \check H ^{\zb} \, = \, (\dab -
H_{\tb} ^{\ \zb} \pab ) \check H ^{\zb} \, = \,
\sum_{n=0}^{\infty}
(- \pab \check H ^{\zb} )^n
\ \dab  \check H ^{\zb}
\ \ .
\end{eqnarray}
The last expression on the r.h.s. of these equations
follows by iteration
of the corresponding equation.
The new variable $H^{\zb}$ ($\check H ^{\zb}$) still allows
for the addition of a superfield
$G^{\zb}$ ($\check G ^{\zb}$) satisfying
$\hat{D} G^{\zb} =0$
$(\hat{\dab} \check G ^{\zb} =0$).
The infinitesimal transformation laws of $H^{\zb}$ and
$\check{H} ^{\zb}$ read
\begin{eqnarray}
s H^{\zb} & = & C^{\zb} ( 1 + \pab H^{\zb} ) + B^{\zb}
\quad , \quad
s B^{\zb} = - C^{\zb} \pab B^{\zb}
\qquad {\rm with} \quad
\hat D B^{\zb} =0 \quad
\nonumber
\\
s \check H ^{\zb} & = & C^{\zb} ( 1 + \pab \check H ^{\zb} ) +
\check B ^{\zb}
\quad , \quad
s \check B^{\zb} = - C^{\zb} \pab \check B^{\zb}
\qquad {\rm with} \quad
\hat{\dab} \check B ^{\zb} =0 \quad
\end{eqnarray}
and induce the transformation laws (\ref{35b}) of
$H_{\th}^{\ \zb}$ and
$H_{\tb}^{\ \zb}$.

We note that the introduction and transformation laws
of $H^{\zb}$ and $\check H ^{\zb}$ are very reminiscent
of the prepotential $V$ occuring in 4-dimensional supersymmetric
Yang-Mills theories: in the abelian case, the latter transforms
according to $sV = i (\LA - \LB)$ where
$\LA$ ($\LB$) represents a chiral (anti-chiral) superfield.

For the restricted geometry, we have $\check H ^{\zb} = 0$ and,
in the WZ-gauge, the non-vanishing component fields
of $H^{\zb}$  and $B ^{\zb}$ are
\[
[ D, \dab ] H^{\zb}  \vert \, =  -2 \bar{\mu}
\qquad {\rm and} \qquad
B^{\zb}  \vert \, =  - \bar c \quad , \quad
[D, \dab ] B^{\zb}  \vert \, =   - (\pa - 2 \bar{\mu} \pab ) \bar c
\ .
\]
In this gauge,
the {\em superdiffeomorphism anomaly in the $\zb$-sector} takes
the form
\begin{equation}
\label{azb}
{\cal A}^{(\zb)} [ C^{\zb} ; H^{\zb} ] \; = \;
\int_{{\bf SC}} d^4 z \, C^{\zb} \pab^3 H^{\zb} \; = \;
- \int_{{\bf C}} d^2 z \, \bar c \, \pab ^3 \bar{\mu}
\ \ .
\end{equation}

\section{Super Beltrami equations}

Substitution of the expressions (\ref{13a}) into the
definitions (\ref{15}) yields the {\em super Beltrami
equations}, e.g. the one involving the basic variable
$\HZB$:
\begin{equation}
0= ( \pab Z + {1 \over 2} \TB \pab \TH + {1 \over 2} \TH \pab \TB )
- \HZB
( \pa Z + {1 \over 2} \TB \pa \TH + {1 \over 2} \TH \pa \TB )
\ \ .
\end{equation}
These equations can be used to define quasi-superconformal
mappings \cite{ta,jc}; they occur in the supergravity approach
\cite{ar} and have been studied from the mathematical
point of view for the $N=1$ case in reference \cite{cr}.

\chapter{(2,2) Theory}

\section{Introduction}

We now summarize the main results
of the (2,2) theory.
As expected,
most expressions in the $z$-sector are the same as those of the
(2,0) theory, while those
in the $\bar{z}$-sector are simply obtained by complex conjugation.
Therefore, our presentation  closely follows the lines of chapter 3
and the new features are pointed
out whenever they show up.
The
general framework for (2,2) SRS's and superconformal transformations
is the one described in chapter 2.

\section{Beltrami superfields}

Starting from a reference complex structure
given by local coordinates
$(z,\th,\tb;\zb,\th^-,\tb^-)$ on a (2,2) SRS, we pass
over to a generic complex structure corresponding to local coordinates
$(Z,\TH,\TB;\ZB,\TH^-,\TB^-)$
by a smooth change of coordinates.
The induced transformation law of the
canonical 1-forms has the form
\begin{equation}
(e^Z,e^{\ZB},e^{\TH},e^{\TB},e^{\TH ^-},e^{\TB ^-})\ =\
(e^z,e^{\zb},e^{\th},e^{\tb},e^{\th ^-},e^{\tb ^-})\ \cdot M \cdot Q
\ \ ,
\end{equation}
where the
matrices $M$ and $Q$ contain the Beltrami superfields and
integrating factors, respectively. More explicitly, $MQ$ reads
\begin{equation}
\left( \begin{array}{llllll}
1        & \Hzzb   & 0       & 0        & \Hztm    & \Hztbm    \\
\Hzbz    & 1       & \Hzbt   & \Hzbtb   & 0        & 0         \\
\Htz     & \Htzb   & \Htt    & \Httb    & \Httm    & \Httbm    \\
\Htbz    & \Htbzb  & \Htbt   & \Htbtb   & \Htbtm   & \Htbtbm   \\
\Htmz    & \Htmzb  & \Htmt   & \Htmtb   & \Htmtm   & \Htmtbm   \\
\Htbmz   & \Htbmzb & \Htbmt  & \Htbmtb  & \Htbmtm  & \Htbmtbm  \\
\end{array}   \right)
\left( \begin{array}{ccllll}
\LA \LB   & 0         &   \tau   &  \taub   & 0         & 0      \\
0         & \LAm \LBm & 0        & 0        & \taum     & \taubm \\
0         & 0         & \LA      & 0        & 0         & 0      \\
0         & 0         & 0        & \LB      & 0         & 0      \\
0         & 0         & 0        & 0        & \LAm      & 0      \\
0         & 0         & 0        & 0        & 0         & \LBm   \\
\end{array}   \right)
,
\end{equation}
where the indices $z, \th, \tb$ and $\zb, \th^- , \tb^-$ are
related by complex conjugation, e.g.
\begin{eqnarray*}
\LA^{\ast} & = & \LA^-
\quad , \quad
\tau^{\ast} \; = \; \tau^-
\quad , \quad
(\Hzbz )^{\ast} \; = \; H_z^{\ \zb}
\quad , \quad
(H_{\tb}^{\ \th} )^{\ast} \; = \;
H_{\tb^-}^{\ \th^-}
\\
\LB^{\ast} & = & \LB^-
\quad , \quad
\bar{\tau} ^{\ast} \; = \; \bar{\tau} ^-
\quad , \quad
(H_{\th}^{\ z})^{\ast} \; = \; H_{\th^-} ^{\ \zb}
\quad , \qquad \qquad \ ...
\end{eqnarray*}
The
`$H$' are invariant under superconformal transformations of the capital
coordinates while the integrating factors change under the latter
according to
\begin{equation}
\begin{array}{cclcccl}
\LA ^{\prime} & = & {\rm e} ^{-W} \ \LA
 & , &
\LB ^{\prime} & = &\ {\rm e} ^{-\bar{W}} \ \LB\\
\tau^{\prime} & = & {\rm e} ^{-W} \ [ \, \tau \ - \
\LA \, \LB \, (D_{\TB} W ) \,]
 & , & \,
\bt^{\prime} & = & {\rm e} ^{-\bar{W}} \ [ \, \bt \ - \
\LA \, \LB \,  (D_{\TH} \bar{W} ) \,]
\ \ ,
\end{array}
\end{equation}
where
${\rm e}^{- W} \equiv D_{\TH} \TH^{\prime}$ and
${\rm e}^{- \bar{W}} \equiv D_{\TB} \TB^{\prime}$.
The transformation laws of $\LA ^-,\LB ^-, \tau ^-,\bt ^-$
are obtained by complex conjugation and involve
$W^{\ast} = W^- , \bar{W} ^{\ast} = \bar W ^-$.

The $U(1)$ symmetry (with parameter $K$)
of the (2,0) theory becomes a $U(1)
\otimes U(1)$-symmetry parametrized by $K$ and $K^- = K^{\ast}$
under which the fields transform according to
\begin{eqnarray}
\LA ^{\prime} & = & {\rm e} ^K \ \LA
\ \ \ \ \ \ \ \ \ , \ \ \ \ \ \ \ \ \ \ \ \
\LB ^{\prime} \ = \ {\rm e} ^{-K} \ \LB
\\
(H_a ^{\ \th} )^{\prime} & = &
{\rm e}^{-K} \
H_a ^{\ \th}
 \ \ \ \ \ ,  \ \ \ \ \ \
(H_a ^{\ \tb} )^{\prime} \ = \
{\rm e}^{K} \
H_a ^{\ \tb}
\ \ \ \ \ {\rm for} \ \ a \, \neq \, z
\nonumber
\end{eqnarray}
and the c.c. equations.

Due to the structure relations (\ref{10}),
the `$H$'
satisfy the following set of equations (and their c.c.):
\begin{eqnarray}
\HO  H_{\tb} ^{\ \tb} \, + \, H_{\tb} ^{\ \th}  H_{\th} ^{\ \tb}
& = & 1\, - \,
(  \dab  -  \HB  \pa  )   \HT  \, - \,
(  D  -  \HT  \pa  )   \HB
\nonumber  \\
\Htmt  H_{\tb ^-} ^{\ \tb} \, + \,
H_{\tb ^-} ^{\ \th}  H_{\th ^-} ^{\ \tb}
& = & \Hzbz\, - \,
(  \dab _-  -  \Htbmz  \pa  )   \Htmz  \, - \,
(  D_-  -  \Htmz  \pa  )   \Htbmz
\label{relH1}\\
H_a ^{\ \th} H_a ^{\ \tb} & = & - \,
(D_a - H_a ^{\ z} \pa ) H_a ^{\ z}
\qquad \qquad \qquad
\qquad {\rm for} \ \; a = \th , \tb , \th^- , \tb^-
\nonumber \\
\Hzbt H_a ^{\ \tb} \, + \, \Hzbtb H_a ^{\ \th}
 & = &
(  D_a  -  H_a ^{\ z}  \pa  ) \HZB \, - \,
(\pab - \HZB \pa ) H_a ^{\ z}
\quad
{\rm for} \ \, a =  \th , \tb , \th ^- , \tb ^-
\nonumber \\
H_a ^{\ \th} H_b ^{\ \tb} +
H_b ^{\ \th} H_a ^{\ \tb} & = & - \,
(D_a - H_a ^{\ z} \pa ) H_b ^{\ z}
\, - \,
(D_b - H_b ^{\ z} \pa ) H_a ^{\ z}
\nonumber  \\
& & \qquad \qquad\quad
{\rm for} \ \, (a,b) =
(\th , \th^-)  , \,
(\th , \tb^-)  , \,
(\tb , \th^-)  , \,
(\tb , \tb^-)
\ .
\nonumber
\end{eqnarray}
By linearizing the variables
($H_{\th}^{\ \th} = 1 + h_{\th} ^{\ \th}, \,
H_{\tb}^{\ \tb} = 1 + h_{\tb} ^{\ \tb}$ and
$H_a ^{\ b} = h_a ^{\ b}$ otherwise), we find that the independent
linearized fields are
$h_{\th}^{\ z}, \,
h_{\tb}^{\ z}, \,
h_{\th}^{\ \th} - h_{\tb}^{\ \tb}, \,
h_{\th^-}^{\ z}, \,
h_{\tb^-}^{\ z}$ where the latter two satisfy (anti-) chirality
conditions
($D_- h_{\th^-}^{\ z} = 0 = \dab_-
h_{\tb^-}^{\ z}$). Thus, there are 5 independent Beltrami superfields,
$H_{\th}^{\ z}, \,
H_{\tb}^{\ z}, \,
H_{\th^-}^{\ z}, \,
H_{\tb^-}^{\ z}$ and
$H_{\th}^{\ \th} / H_{\tb}^{\ \tb}$, but
$H_{\th^-}^{\ z}$ and
$H_{\tb^-}^{\ z}$ satisfy chirality-type conditions
which reduce the number of their independent component fields
by a factor 1/2.
In section 4.8, these constraints will be explicitly solved in a
special case in terms of an unconstrained superfield $H^z$.

The factors $\tau , \, \bar{\tau}$ are differential polynomials
of the Beltrami coefficients and of the
integrating factors $\LA , \LB$:
\begin{eqnarray}
\tau & = &
( \HO  H_{\tb} ^{\ \tb}  +  H_{\tb} ^{\ \th}  H_{\th} ^{\ \tb}
 )^{-1}  \left[  (  \dab  -  \HB  \pa  )
( \HO  \LA  ) +
(  D  -  \HT  \pa  )
( H_{\tb} ^{\ \th}  \LA  )   \right] \ \\
\bar{\tau} & = &
( \HO  H_{\tb} ^{\ \tb}  +  H_{\tb} ^{\ \th}  H_{\th} ^{\ \tb}
 )^{-1}  \left[  (  D  -  \HT  \pa  )
( H_{\tb} ^{\ \tb}  \LB  ) +
(  \dab  -  \HB  \pa  )
( H_{\th} ^{\ \tb}  \LB  )   \right]
\ .
\nonumber
\end{eqnarray}
As for the factors
$\LA , \LB$ themselves, they satisfy the IFEQ's
\begin{eqnarray}
0 & = &
(\, D_a  -  H_a ^{\ z} \pa   - \frac{1}{2} \, \pa H_a ^{\ z}
 -  V_a ) \, \LA
\,  - \, H_a ^{\ \tb} \, \tau
\\
0 & = & (\, D_a  - H_a ^{\ z}
\pa   - \frac{1}{2} \, \pa H_a ^{\ z}
 +  V_a ) \, \LB
\,  - \, H_a ^{\ \th} \, \bt
\ \ ,
\nonumber
\end{eqnarray}
where it is understood
that $H_z^{\ z} =1$ and $H_z^{\ \th} = 0 = H_z^{\ \tb}$.
The c.c. variables $\LA^- , \bar{\LA} ^- , \tau^- , \bar{\tau} ^-$
satisfy the c.c. equations and
the $U(1) \otimes U(1)$ connections $V_a$ and $V^- _a$
which appear in the previous set of equations
are given by
\begin{eqnarray}
V_z & = & 0 \nonumber \\
V_{\zb} & = & \frac{1}{\Htt} \{[ D - \Htz \pa + \frac{1}{2}
(\pa \Htz) + V_{\th}] \, H_{\zb} ^{\ \th}
\, - \, [ \pab - \Hzbz \pa + \frac{1}{2}
(\pa \Hzbz)] \, \Htt \} \nonumber \\
V_{\th} & = & - \frac{1}{\Htt}\ [ D - \Htz \pa + \frac{1}{2}
(\pa \Htz) ] \, \Htt \\
V_{\tb} & = &  \frac{1}{\Htbtb}\ [ \bar{D} - \Htbz \pa + \frac{1}{2}
(\pa \Htbz ) ] \, \Htbtb \nonumber \\
V_a & = & -\frac{1}{\Htt} \{ [ D_a - H_a ^{\ z} \pa + \frac{1}{2}
(\pa H_a ^{\ z}) ] \, \Htt \, +\,
[ D - \Htz \pa + \frac{1}{2} (\pa \Htz) + V_{\th}] \, H_a ^{\ \th} \}
\nonumber \\
&& \qquad \qquad \qquad \qquad \qquad \qquad
\qquad \qquad \qquad \qquad \qquad
\qquad {\rm for} \ \, a = \th ^- , \tb ^- \ . \nonumber
\end{eqnarray}
We note that
the last equations can also be written in the form
\begin{eqnarray}
H_a ^{\ \th} V_a & = & \ -[ D_a - H_a ^{\ z} \pa + \frac{1}{2} (\pa H_a
^{\ z}) ]
\, H_a ^{\ \th} \ \, \ \ \ {\rm for} \ \, a= \tb , \th ^- , \tb ^-
\nonumber \\
H_a ^{\ \tb} V_a & = & \ [ D_a - H_a ^{\ z} \pa + \frac{1}{2} (\pa H_a
^{\ z}) ] \, H_a ^{\ \tb}
\qquad \ {\rm for} \ \, a = \th , \th ^- , \tb ^- \
 .
\end{eqnarray}

\section{Symmetry transformations}

In order to obtain the
transformation laws of the fields
under infinitesimal superdiffeomorphisms
and $U(1) \otimes U(1)$ transformations,
we introduce the ghost vector field
\[
\Xi \cdot \pa \ \equiv \
\Xi^{z} \, \pa \ + \
\Xi^{\zb} \, \pab \ + \
\Xi^{\th} \, D \ + \
\Xi^{\tb} \, \dab \ + \
\Xi^{\th ^-} \, D_- \ + \
\Xi^{\tb ^-} \, \dab _-
\ \ ,
\]
(with $\Xi^a =\, \Xi^a (z , \th , \tb \, ; \zb , \th ^- , \tb ^- )$)
which generates an infinitesimal change of the coordinates
$(z , \th , \tb \, ; \zb , \th ^- , \tb ^-)$.

The $U(1) \otimes U(1)$ transformations again appear
in a natural way in
the trans\-formation laws of the integrating factors
and are parametrized by
ghost superfields
$K$ and $K ^-$ .
In terms of the reparametrized ghosts
\begin{equation}
\left( \,
C^z \, ,\, C^{\zb}  \, ,\,
C^{\th} \, ,\, C^{\tb}\, ,\, C^{\th ^-} \, ,\, C^{\tb ^-}
\right) \ = \
\left( \, \Xi^z \, , \, \Xi^{\zb} \, , \,
\Xi ^{\th} \, ,\, \Xi ^{\tb} \, , \,
\Xi ^{\th ^-} \, ,\, \Xi ^{\tb ^-}
\, \right) \cdot M
\ \ ,
\end{equation}
the BRS variations read
\begin{eqnarray}
s \LA & = & C^z \, \pa \LA  \ + \ \frac{1}{2} \  (\pa C^z) \, \LA
\ + \  C^{\tb} \, \tau \ + \ K\, \LA
\nonumber \\
s \LB & = & C^z \, \pa \LB  \ + \ \frac{1}{2} \  (\pa C^z) \, \LB
\ + \  C^{\th} \, \bt \ - \ K\, \LB
\nonumber \\
s \tau & = &  \pa \, ( \, C^z  \tau  +   C^{\th}
\LA \, )
\\
s \bt & = &  \pa  \, ( \, C^z  \bt  +   C^{\tb}
\LB \, )
\ \ ,
\nonumber
\end{eqnarray}
\begin{eqnarray}
\label{ul}
sH_a ^{\ z} & = &
(\, D_a   -  H_a ^{\ z} \pa
 +    \pa H_a ^{\ z}  \, )\, C^z  -  H_a ^{\ \th}
C^{\tb}  -  H_a ^{\ \tb}
C^{\th}
\\
s H_a ^{\ \th} & = & ( \, D_a  -  H_a ^{\ z}   \pa
+  \frac{1}{2} \,  \pa H_a ^{\ z}  +  V_a \, ) \, C^{\th}
 +  C^z   \pa H_a ^{\ \th}   -
\frac{1}{2} \,
H_a ^{\ \th}  ( \pa C^z )   -  H_a ^{\ \th}  K
\nonumber \\
s H_a ^{\ \tb} & = & ( \, D_a  -  H_a ^{\ z}   \pa
+  \frac{1}{2} \,  \pa H_a ^{\ z}  -  V_a \, )  \, C^{\tb}
+  C^z   \pa H_a ^{\ \tb}   -
\frac{1}{2} \,
H_a ^{\ \tb}  ( \pa C^z )  +  H_a ^{\ \tb}  K
\nonumber
\\
sV_a &  =  & C^z  \pa V_a  +  \frac{1}{2}  H_a ^{\ \th}
\pa C^{\tb}
 -  \frac{1}{2}  (\pa H_a ^{\ \th}) C^{\tb}
 -  \frac{1}{2}  H_a ^{\ \tb}
\pa C^{\th}
+ \frac{1}{2}  (\pa H_a ^{\ \tb} ) C^{\th}
\nonumber \\
& & \qquad \qquad \qquad \qquad \qquad
\qquad \qquad \qquad \qquad \qquad
\quad
+ ( D_a  -  H_a ^{\ z}  \pa  )  K
\nonumber
\end{eqnarray}
\begin{eqnarray}
s C^z & = & -  \, [ \, C^z  \pa C^z  +  C^{\th}
C^{\tb} \, ]
\nonumber
\\
s C^{\th} & = & - \, [ \, C^z  \pa  C^{\th}  +  \frac{1}{2}
\, C^{\th}  (\pa C^z )  -  K  C^{\th}  \, ]
\nonumber
\\
\label{ult}
s C^{\tb} & = & - \, [ \, C^z  \pa  C^{\tb}  +  \frac{1}{2}
\, C^{\tb}  (\pa C^z )  +  K  C^{\tb}  \, ]
\\
s K & = & - \, [ \, C^z  \pa  K  -   \frac{1}{2} \,
 C^{\th}  (\pa C^{\tb} ) +  \frac{1}{2} \, C^{\tb}
(\pa C^{\th} ) \, ]
\ \ .
\nonumber
\end{eqnarray}
The variations of the c.c. fields are simply obtained
by complex conjugation and henceforth
the holomorphic factorization is manifestly
realized for the chosen parametrization.
Furthermore, the number of independent Beltrami fields and
the number of
symmetry parameters coincide. By
projecting to space-time fields according to eqs.(\ref{39})(\ref{40}),
one obtains the transformation laws
(\ref{41}).
The variations (\ref{ul})(\ref{ult}) of $H_a^{\ b}, \, V_a ,\, C^a$
and $K$ coincide with
those found in the metric approach in reference \cite{ot}.

\section{Scalar superfields}

We consider complex
superfields
${\cal X}^i$ and $\bar{{\cal X}} ^{\bar{\imath}} =
({\cal X}^i)^{\ast}$
satisfying the (twisted) chirality conditions \cite{ghr}
\begin{equation}
\begin{array}{rcl}
D_{\TB} {\cal X}^i  &  = &  0  \  =  \
D_{\TH^-} {\cal X}^i
\\
D_{\TH} \bar{{\cal X}} ^{\bar{\imath}} &  =  &  0  \  =  \
D_{\TB ^-} \bar{{\cal X}} ^{\bar{\imath}}
\ \ .
\end{array}
\end{equation}
Other multiplets have been introduced and discussed in references
\cite{ghr} and \cite{ggw}.
The sigma-model action
describing the coupling of these fields to a superconformal class
of metrics on the SRS ${\bf {S\Sigma}}$ is given by
\cite{bz,ghr}
\begin{equation}
S_{inv} [ {\cal X} ,  \bar{{\cal X}} ] \, =\int_{\bf {S\Sigma}}d^6 Z
\ K({\cal X}  ,  \bar{{\cal X}})
\ \ ,
\label{action22}
\end{equation}
where
$K$ is a real function of the fields ${\cal X}\, , \, \bar{{\cal X}}$
and
$d^6 Z = dZ \, d\bar{Z} \, d\Theta \, d\bar{\Theta}
\, d\Theta ^- \, d\bar{\Theta} ^- $ is the
superconformally invariant measure.
For a
flat target space metric, the functional (\ref{action22}) reduces to
\cite{ade}
\begin{equation}
S_{inv} [ {\cal X} ,  \bar{{\cal X}} ] \, =\int_{\bf {S\Sigma}}d^6 Z
\ {\cal X} \bar{{\cal X}}
\ \ .
\label{action22flat}
\end{equation}

\section{Restriction of the geometry}

The restriction of the geometry
is achieved by imposing the following conditions:
\begin{equation}
\label{impo}
\Htz \,  = \, \Htbz \, = \,
\Htbmz \, = \, 0
\qquad {\rm and} \qquad
\Htt / \Htbtb \, = \, 1
\ \ .
\end{equation}
The addition of the
c.c. equations goes without saying in this whole section.
Equations (\ref{relH1}) then imply that
all Beltrami coefficients depend on $\Htmz$ by virtue of the relations
\begin{eqnarray}
\Hzbz & = &  \bar{D}_- \Htmz
\quad , \quad
\Hzbt \, = \,  \bar{D} \Hzbz
\quad , \quad
\Htmt \, = \, - \bar{D} \Htmz
\nonumber
\\
& & \qquad \qquad \quad\ \
\Hzbtb \, = \, D \Hzbz
\quad , \quad
\Htmtb \, = \, - D\Htmz
\\
H_{\th}^{\ \tb} & = &
H_{\tb}^{\ \th} \, = \,
\Htbmt \, =  \,  \Htbmtb \, = \, 0
\ \ \quad , \quad
\Htt \, = \, 1 \, = \, \Htbtb
\nonumber
\end{eqnarray}
and that $\Htmz$ itself satisfies the covariant chirality condition
\begin{equation}
\label{grig}
(D_- - \Htmz \pa
 + D\Htmz \, \dab ) \, \Htmz \,  = \, 0
\ \ .
\end{equation}
The relations satisfied by the other variables become
\begin{eqnarray}
\tau & = & \bar{D} \LA
\quad , \quad
D \LA \; = \; 0
\quad , \quad
\bar{D}_- \LA \; = \;  0
\quad , \quad
D_- \LA \,=\, D\bar{D}(\Htmz \LA )
\nonumber
\\
\bar{\tau} & = & D \LB
\quad , \quad
\bar{D} \bar{\LA} \; = \; 0
\quad , \quad
\bar{D}_- \bar{\LA} \; = \; 0
\quad , \quad
D_- \LB \,=\, \bar{D}D(\Htmz \LB )
\nonumber
\\
V_{\th} & = & 0
\quad \quad , \quad \
V_{\thm} \;=\; \frac{1}{2} \, [D,\bar{D}] \Htmz
\quad \qquad , \qquad
V_{\zb} \;=\; \bar{D}_- V_{\thm}
\\
V_{\tb} & = & 0
\ \ \ \quad , \ \quad
V_{\tbm}\;=\; 0
\quad .
\nonumber
\end{eqnarray}
and eqs.(\ref{13a})(\ref{ana}) yield the local expressions
\begin{equation}
\LA \,=\,D\TH \ \ \ ,\ \ \ \LB \,=\,\bar{D} \bar{\TH} \ \ .
\end{equation}
The $s$-invariance of conditions (\ref{impo}) implies that the
symmetry parameters $C^{\th}, \, C^{\tb}$ and $K$ depend on $C^z$
according to
\begin{equation}
\begin{array}{lclcl}
C^{\tb} \,=\, DC^z &,& C^{\th} \,=\, \bar{D} C^z &,&
K\,=\,\ds{1 \over 2} \, [ D , \bar{D} ] C^z
\end{array}
\end{equation}
and that $C^z$ itself satisfies the chirality condition
\begin{equation}
\bar{D}_- C^z \,=\, 0 \ \ .
\label{contC}
\end{equation}
Thus, the $s$-variations of the basic variables read
\begin{eqnarray}
s \Htmz & = & [ D_- - \Htmz \pa + (\dab \Htmz ) D + (D \Htmz ) \dab
+ (\pa \Htmz ) ] \, C^z
\nonumber  \\
sC^z & = & - \, [ C^z \pa C^z + (D C^z ) (\dab C^z ) ]
\ \ .
\end{eqnarray}

\section{Intermediate coordinates}

The intermediate coordinates which are relevant for us are those
obtained by going over from $z$ and $\tb$ to capital coordinates
without modifying the other coordinates:
\begin{equation}
(z , \th , \tb \, ; \zb , \th^- , \tb^- )
\ \stackrel{M_1 Q_1}{\longrightarrow} \
(\zt, \tilde{\th} , \tilde{\tb} \, ;
\tilde{\zb} , \tilde{\th}^-  , \tilde{\bar{\th}} ^- )
\equiv
(Z,  \th , \TB \, ; \zb , \th^- , \tb^- )
\ \ .
\end{equation}
For the restricted geometry, we then get the explicit expression
\begin{equation}
\tilde D _- = D_- - \Htmz \pa + (D\Htmz ) \dab
\end{equation}
and by construction we have $(\tilde D _-)^2 =0$. Thus, the covariant
chirality condition (\ref{grig}) for $\Htmz$ reads
$\tilde D _- \Htmz =0$ and may be solved by virtue of the nilpotency
of the operator
$\tilde D _-$ (see section 4.8).

\section{Component field expressions}

To write the action
(\ref{action22}) in terms of the reference coordinates
$(z ,\th ,\tb \, ; \zb,\thm ,\tbm )$,
we introduce the following superfields
(as in the $(2,0)$ case):
\begin{equation}
\begin{array}{lllclll}
\nonumber
h_a^{\ z} &=& \Delta ^{-1} (H_a^{\ z}-\Hzbz H_a^{\ \zb}) &  , &
h_a^{\ \zb} &=& \Delta ^{-1} (H_a^{\ \zb}-\Hzzb H_a^{\ z})\\
h_a^{\ \th} &=& H_a^{\ \th}-h_a^{\ \zb}\Hzbt & , &
h_a^{\ \thm} &=& H_a^{\ \thm}-h_a^{\ z}\Hztm\\
\nonumber
h_a^{\ \tb} &=& H_a^{\ \tb}-h_a^{\ \zb}\Hzbtb & , &
h_a^{\ \tbm} &=& H_a^{\ \tbm}-h_a^{\ z}\Hztbm
\end{array}
\end{equation}
for $a=\th,\thm,\tb,\tbm $ .
In the
remainder of this section, we will consider the restricted geometry,
for which the Berezinian takes the form
\begin{equation}
\left|
\frac{\pa (Z,\TH ,\TB \,; \ZB \,\TH ^- ,\TB ^- )}{\pa (z ,\th ,\tb \, ;
\zb , \thm , \tbm )}
\right|
\ =\ \Delta / h
\end{equation}
with $\Delta = 1-\Hzbz \Hzzb $ and  $h=\htt \htmtm - \htmt \httm $ .
The chirality
conditions for the matter superfields read $\dab {\cal X} =0$ and
\begin{equation}
\htt (D_- - \htmzb \pab - \htmz \pa - \htmtbm \dab_- ) {\cal X} =
\htmt (D -\htzb \pab - \htz \pa -\httbm \dab_-) {\cal X}
\end{equation}
and c.c.  .

We now choose a {\em WZ-gauge} in which
the basic superfields have the $\th$-expansions
\begin{eqnarray}
\label{expa}
\Htmz &=& \tbm (\mu + \tb \alpha + \th \bar{\alpha} + \tb \th \bar{v})
\qquad \qquad , \quad
C^z = c + \tb \epsilon + \th \bar{\epsilon} + \tb \th k \\
\Htzb
&=& \tb (\mb + \tbm \alpha^- + \thm \bar{\alpha}^- + \tbm \thm \bar{v}^-
 ) \ , \quad
C^{\zb}
= \bar{c} + \tbm \epsilon^- + \thm \bar{\epsilon}^- + \tbm \thm   k^-
,
\nonumber
\end{eqnarray}
whose form and physical interpretation is similar to the one of
expressions
(\ref{cf}) of the (2,0) theory.
In fact, we have
$\Htmz = \tbm {\cal H}_{\zb}^{\ z}$ where
${\cal H}_{\zb}^{\ z}$ denotes the basic Beltrami superfield
of the (2,0) theory: a similar relationship holds in the
WZ-gauge between the basic Beltrami superfields
of the (1,1) and (1,0) theories \cite{dg}.

The (twisted chiral)
matter superfields ${\cal X}$ and $\bar{{\cal X}}$
contain one complex scalar, four spinors
and one complex auxiliary fields as component fields \cite{ghr,ggw},
\begin{equation}
\begin{array}{lllllll}
X={\cal X} \! \mid &,& \ \la_{\th}=D{\cal X} \! \mid &,&
\bar{\la}^-_{\tbm}
=\dab_-{\cal X} \! \mid &,&F_{\th \tbm}=D\dab_-{\cal X} \!
\mid
\\
& & & & & &
\\
\bar X
=\bar{{\cal X}} \! \mid &,&\la^-_{\thm}=D_- \bar{{\cal X}} \! \mid &,&
\ \bar{\la}_{\tb}
=\dab \bar{{\cal X}} \! \mid &,&\bar{F}_{\thm \tb}=D_-\dab
\bar{{\cal X}} \! \mid
\end{array}
\end{equation}
for which fields the
action (\ref{action22flat}) reduces to the following functional
on the Riemann surface ${\bf \Sigma}$:
\begin{eqnarray}
\label{n2a}
S_{inv} & = & \int_{\bf \Sigma}  d^2z  \left\{
\ds \frac{1}{1-\mu\mb}  \ [\
(\pa -\mb \pab) \bar{X} \, (\pab -\mu \pa)X
\ - \ \alpha \la (\pa -\mb \pab)\bar{X} \right.
\\
&& \qquad \qquad - \ \alpha^-\la^-(\pab -\mu \pa)X
\ - \ \bar{\alpha}\lb (\pa -\mb \pab)X
\ - \ \bar{\alpha}^-\lb^- (\pab -\mu \pa)
\bar X
\nonumber
 \\
&& \qquad \qquad + \ (\alpha\la)(\alpha^-\la^--\mb\bar{\alpha}\lb)
\ + \ (\bar{\alpha}^-\lb^-)(\bar{\alpha}\lb-\mu\alpha^-\la^-)\ ]
\nonumber
\\
&& \qquad \qquad -\ \lb (\pab -\mu \pa -\frac{1}{2}\pa\mu -\bar{v})\la
\ - \ \lb^- (\pa -\mb \pab -\frac{1}{2}\pab\mb -\bar{v}^-)\la^-
\nonumber
\\
&& \qquad \qquad \left. - \ (1-\mu\mb)\bar{F} F \right\} \ .
\nonumber
\end{eqnarray}
In terms of
$\xi^a = \Xi^a \! \mid$ and the short-hand notation
\begin{eqnarray*}
\xi  & \equiv & \xi^z \quad , \quad
\hat{k} \equiv k- \bar{\xi} \bar{v} \quad , \quad
\xi \cdot \pa \equiv \xi \pa + \bar{\xi}
\pab
\\
\bar{\xi} & \equiv & \xi^{\zb} \quad , \quad
\hat{k}^-\equiv k^--\xi \bar{v}^-
\ \ ,
\end{eqnarray*}
the $s$-variations of the matter fields read
\begin{eqnarray}
\nonumber
sX &=&(\xi \cdot \pa )X+\xi^{\th} \la + \xi^{\tbm} \bar{\la}^- \\
\nonumber
s \la
&=& [(\xi \cdot \pa )+\frac{1}{2} (\pa \xi + \mu \pa \bar{\xi} ) +\hat{k}
 ] \, \la \ + \
\xi^{\tb}{\cal D}X\ -\
\xi^{\tbm} F \\
\label{sba}
s\bar{\la}^-
&=& [(\xi \cdot \pa )+\frac{1}{2} (\pab \bar{\xi} + \mb \pab \xi
 ) -\hat{k}^- ]
\, \bar{\la}^- \ +\
\xi^{\thm}\bar{{\cal D}}X\ +\
\xi^{\th} F \\
\nonumber
sF &=& [(\xi \cdot \pa )+\frac{1}{2} (\pa \xi + \mu \pa \bar{\xi} ) +
\frac{1}{2} (\pab \bar{\xi} + \mb \pab \xi ) +\hat{k}-\hat{k}^- ] \, F
\\
&& \qquad \qquad \qquad \qquad
\ +\ \xi^{\tb} {\cal D}\bar{\la}^- \ -\
\xi^{\thm}\bar{{\cal D}}\la \ ,
\nonumber
\end{eqnarray}
where we have introduced
the supercovariant derivatives
\begin{eqnarray}
\nonumber
{\cal D}X &=&
\frac{1}{1-\mu \mb } \ [(\pa -\mb \pab )X+\mb \alpha \la
 -\bar{\alpha}^-\bar{\la}^- ]  \\
\label{sgra}
\bar{{\cal D}}X &=&
\frac{1}{1-\mu \mb }
\ [(\pab -\mu \pa )X+\mu \bar{\alpha}^- \bar{\la}^- -\alpha
 \la ]   \\
{\cal D}\bar{\la}^- &=&
\frac{1}{1-\mu \mb }
\ [(\pa -\mb \pab -\frac{1}{2}\pab \mb +\bar{v}^-)\bar{\la}^-
 +\mb \alpha F -\alpha^- \bar{{\cal D}}X ]
\nonumber \\
\bar{{\cal D}}\la &=&
\frac{1}{1-\mu \mb }
\ [(\pab -\mu \pa -\frac{1}{2}\pa \mu -\bar{v})\la -\mu
\bar{\alpha}^- F-\bar{\alpha} {\cal D}X ]\ .
\nonumber
\end{eqnarray}
A generic expression for the variations of the matter fields
and for the supercovariant derivatives
can be given in the supergravity framework where the component
fields are defined by covariant projection \cite{ot}.
We leave it as an exercise to check that the action (\ref{n2a})
describing the superconformally invariant coupling of a twisted chiral
multiplet to supergravity coincides with the usual component
field expression \cite{ggw} by virtue of the Beltrami parametrization
of the space-time gauge fields (i.e. the zweibein, gravitino
and $U(1)$ gauge field) - see \cite{bbgc, gg} for the $N=1$ theory.
Component field results for a chiral multiplet can be
directly obtained from our results for
the twisted chiral multiplet by application of the
mirror map \cite{ggw}.

\section{Anomaly}

As pointed out in section 4.6, the constraint satisfied by $\Htmz$
in the restricted geometry, i.e.
$\tilde D _- \Htmz =0$, can be solved by virtue of the nilpotency
of the operator
$\tilde D _-$:
\begin{equation}
\Htmz = \tilde D _- H^z =
[D_- - \Htmz \pa + (D\Htmz ) \dab] \, H^z
\ \ .
\end{equation}
Here, the new variable
$H^z$ is determined up to a superfield $G^z$ satisfying
$\tilde D_- G^z =0$ and it transforms according to
\begin{eqnarray}
sH^z & = & C^z \, (1 + \pa H^z ) + (DC^z)(\dab H^z) + B^z
\qquad  {\rm with} \quad
\tilde{D} _- B^z = 0
\nonumber \\
sB^z & = & - \, [ C^z \pa B^z  + (DC^z)(\dab B^z) ]
\ \ .
\label{true}
\end{eqnarray}
In the WZ-gauge, we have $H^z = \th^- H_{\th^-}^{\ z}$
with $H_{\th^-}^{\ z}$ given by (\ref{expa}). In this case, the
{\em holomorphically split form of the superdiffeomorphism anomaly}
on the superplane reads
\begin{eqnarray}
\label{438}
{\cal A} [C^z ; H^z ] \ + \ {\rm c.c.} & = &
\int_{\bf SC} d^6z \ C^z \, \pa [ D, \dab ] \, H^z
\ + \ {\rm c.c.}
\\
& = &
- {1 \over 2 } \
\int_{\bf C} d^2z \ \left\{ c \pa^3 \mu
+ 2 \epsilon \pa^2 \bar{\alpha}
+ 2 \bar{\epsilon} \pa^2 \alpha
+ 4 k \pa \bar v \right\}
\ + \ {\rm c.c.}
\ \ .
\nonumber
\end{eqnarray}
It satisfies the consistency condition
$s{\cal A} = 0$ and can be generalized
to a generic compact SRS
by replacing the operator
$\pa [D, \dab ]$ by the superconformally covariant operator (\ref{bol}).
The component field expression (\ref{438}) coincides with the one found
for the $z$-sector of the (2,0) theory, eq.(\ref{an}), and
with the one of references \cite{y} and \cite{ot} where
other arguments have been invoked.

At the linearized level, the transformation law
(\ref{true}) of $H^z$ reads
\[
\delta H^z  =  C^z + B^z
\qquad  {\rm with} \quad
\bar{D} _- C^z = 0=
D_- B^z
\ \ .
\]
By solving the given constraints on $C^z$ and $B^z$ in terms of
spinorial superfields $L^{\th}$ and $L^{\prime \tb}$,
one finds
\begin{equation}
\delta H^z  =  \bar{D} _- L^{\th} + D_- L^{\prime \tb}
\ \ ,
\end{equation}
which result has the same form as the one found in the second
of references \cite{gw}, see eq.(3.19).

\chapter{Conclusion}

In the course of the completion of our
manuscript\footnote{A
preliminary version of the present paper has been part of the
habilitation thesis of F.G. (Universit\'e de Chamb\'ery, December
1994).},
the work \cite{l} concerning the
(2,0) theory appeared which also discusses the generalization
of our previous $N=1$ results
\cite{dg, dg1}. However, the author of reference \cite{l}
fails to take properly into account the U(1)-symmetry,
connection and transformation laws which leads to incorrect results
and conclusions. Furthermore, the super Beltrami coefficients
(2.34) of \cite{l} are not inert under superconformal transformations
of the capital coordinates, eqs.(2.33), and therefore do not
parametrize superconformal structures as they are supposed to.
Finally, various aspects of the (2,0) theory that we treat here
(e.g. superconformal models and
component field expressions)
are not addressed in reference \cite{l}.

In a supergravity approach \cite{ggrs} , some gauge choices
are usually made when an explicit
solution of the constraints is determined. Therefore, the question
arises in which case the final solution represents
a complete solution of the problem, i.e.
a complete set of prepotentials (and compensators).
Obviously, such a solution has been obtained if
there are as many independent variables as there are independent
symmetry parameters in the theory. If there is
a smaller number of prepotentials, then it
is clear that some basic symmetry parameters have been used to
eliminate fields from the theory (a `gauge choice' or
`restriction of the geometry' has been made).
From these facts, we conclude that
the solution of constraints discussed in references \cite{eo,
kl, l} and \cite{gw} is not complete.
As for
reference \cite{ot}, it has not been investigated which ones
are the independent variables.

Possible
further developments or applications of our formalism include
the derivation of operator product expansions
and the proof of
holomorphic factorization of partition functions
along the lines of the work on the $N=1$ theory
\cite{dg1, agn}. (The latter reference also involves the
supersymmetric generalization of the Verlinde functional
which occurs in conformal field theories and in the
theory of $W$-algebras.)
Another extension of the present study consists of the determination
of $N=2$ superconformally covariant differential operators
and of their application to super $W$-algebras.
This development will be reported on elsewhere \cite{ip}.

\nopagebreak

\end{document}